\documentclass[12pt,preprint]{aastex}
\usepackage{emulateapj5}           
\newcommand{\gsim}{\raisebox{-0.3ex}{\mbox{$\stackrel{>}{_\sim} \,$}}}

\slugcomment{Version from \today}

\begin{document}

\title{High-Mass Proto-Stellar Candidates - \\ I : The Sample and Initial Results}
\author{T.K.Sridharan}
\email{tksridha@cfa.harvard.edu}
\affil{Harvard-Smithsonian Center for Astrophysics, 60 Garden Street, MS 78, Cambridge, MA 02138, USA.}
\author{H. Beuther, P. Schilke, K.M. Menten}
\email{beuther@mpifr-bonn.mpg.de, schilke@mpifr-bonn.mpg.de, kmenten@mpifr-bonn.mpg.de}
\affil{Max-Planck-Institut f\"ur Radioastronomie, Auf dem H\"ugel 69, 53121 Bonn, Germany}
\author{F. Wyrowski}
\email{wyrowski@astro.umd.edu}
\affil{Department of Astronomy, University of Maryland, College Park, USA}

\begin{abstract}
We describe a systematic program aimed at identifying and
characterizing candidate high-mass proto-stellar objects (HMPOs). Our
candidate sample consists of 69 objects selected by criteria based on
those established by Ramesh \& Sridharan (1997) using far-infrared,
radio-continuum and molecular line
data. Infrared-Astronomical-Satellite ({\it IRAS}) and
Midcourse-Space-Experiment (MSX) data were used to study the larger
scale environments of the candidate sources and to determine their
total luminosities and dust temperatures.

To derive the physical and chemical properties of our target regions,
we observed continuum and spectral line radiation at millimeter
and radio wavelengths. We imaged the free-free and dust continuum
emission at wavelengths of 3.6~cm and 1.2~mm, respectively, searched
for H$_2$O and CH$_3$OH maser emission and observed the CO $J=2\to1$
and several NH$_3$ lines toward all sources in our sample. Other
molecular tracers were observed in a subsample.

While dust continuum emission was detected in all sources, most of
them show only weak or no emission at 3.6~cm. Where detected, the cm
emission is frequently found to be offset from the mm emission,
indicating that the free-free and dust emissions arise from different
subsources possibly belonging to the same (proto)cluster. A comparison
of the luminosities derived from the cm emission with bolometric
luminosities calculated from the {\it IRAS} far-infrared fluxes shows
that the cm emission very likely traces the most massive source,
whereas the whole cluster contributes to the far-infrared
luminosity. Estimates of the accretion luminosity indicate that a
significant fraction of the bolometric luminosity is still due to
accretion processes. The earliest stages of HMPO evolution we seek to
identify are represented by dust cores without radio emission.

Line wings due to outflow activity are nearly omnipresent in the CO
observations, and the molecular line data indicate the presence of hot
cores for several sources, where the abundances of various molecular
species are elevated due to evaporation of icy grain
mantles. Kinetic gas temperatures of 40 sources are derived from NH$_3$
(1,1) and (2,2) data, and we compare the results with the
dust temperatures obtained from the {\it IRAS} data.

Comparing the amount of dust, and hence the gas, associated with the
HMPOs and with ultracompact H{\sc ii} regions (UCH{\sc ii}s) we find
that the two types of sources are clearly separated in mass-luminosity
diagrams: for the same dust masses the UCH{\sc ii} regions have higher
bolometric luminosities than HMPOs. We suggest that this is an
evolutionary trend with the HMPOs being younger and reprocessing less
(stellar) radiation in the IR than the more evolved UCH{\sc ii}s
regions.

These results indicate that a substantial fraction of our sample
harbors HMPOs in a pre-UCH{\sc ii} region phase, the earliest known
stage in the high-mass star formation process. 

\end{abstract}


\section{Introduction}

High-mass stars exert a decisive influence on the appearance and
evolution of galaxies. Throughout their life cycle, they inject
significant amounts of energy and momentum into their environments
through stellar winds, molecular outflows, ultraviolet (UV) radiation
and eventually supernova explosions. They also influence the formation
of nearby low-mass stars, most of which are formed in dense clusters
heavily affected by a few massive stars, the Trapezium cluster in
Orion being a well studied nearby example \citep{hillenbrand
1998}. Clearly, questions concerning the number, birth-rates,
distribution, and evolutionary timescales of massive stars are of
critical importance for building a general picture of Galactic
evolution, in addition to their being interesting in themselves. While a
scenario exists for the formation of low mass stars, much less is
known about the processes involved in the formation of high-mass
stars, which we take to be stars with M$\geq 8$M$_{\odot}$
(corresponding to stars of spectral type earlier than B2) not having
an optically visible pre-main-sequence phase
\citep{palla 1993}. This lack of knowledge is due to the fact that it
is difficult to identify the earliest stages of the high-mass star
formation process. Owing to their smaller numbers and shorter
evolutionary timescales, high-mass star-forming regions are on the
average more distant than low mass star-forming regions. Additionally,
the clustered mode in which massive star formation generally seems to
proceed makes it very difficult to resolve and locate young high-mass
protostars, which are deeply embedded and usually show little or no
near-infrared emission (see Menten \& Reid 1995 for the case of the
BN/KL region in Orion).

Recently, much attention has been dedicated to studies of massive
star formation and significant progress is being made. A
handful of sources likely to be young high-mass stellar objects
have been identified. Among the best examples are G31.41+0.31mm
\citep{cesaroni 1994}, G192.16 \citep{shepherd 1998}, {\it IRAS} 20126+4104 
\citep{cesaroni 1997}, {\it IRAS} 23385+6053 \citep{molinari 1998b} and
G34.24+0.13mm \citep{hunter 1998}. 
These objects are characterized by high luminosities
($>10^4$~L$_{\sun}$), dense ($>10^6$~cm$^{-3}$) and warm ($>100$~K)
molecular gas and strong dust emission. Since they only show very weak
or no free-free emission at cm wavelengths from ionized gas, they have
not yet developed an ultracompact H{\sc ii} (UCH{\sc ii}) region.
Henceforth, we will call these objects in a pre-UCH{\sc ii} region
phase high-mass proto-stellar objects, HMPOs for short. In this sense
internally heated hot cores are a sub-class of HMPOs shortly before
ultracompact H{\sc ii} regions become detectable.

The first cases of known HMPOs are in the vicinity of UCH{\sc ii}
regions, a selection effect because they were mostly discovered during
molecular line and/or dust continuum studies of the UCH{\sc ii}
regions.  However, the nearness of an UCH{\sc ii} region (i.e. within,
say, less than 0.1~pc or a few arcsec in angular units) is obviously a
draw-back if one wants to study the earliest and unimpaired stages of
massive star formation: the physical and chemical conditions in the
vicinity of the HMPOs are significantly modified by interactions with
the UCH{\sc ii} regions and their precursors. Also, frequently it is
observationally difficult to discriminate between material associated
with the UCH{\sc ii} region and the nearby HMPO. The latter point can
be partially resolved by high resolution and high dynamic range
interferometric observations, which in some of the cases cited above
have led to the detection of HMPOs in the same fields as the UCH{\sc
ii} regions.

Systematic studies of a substantial sample of isolated HMPOs are
highly desirable and samples of HMPO candidates have been
investigated, e.g. by \citet{molinari 1996} and \citet{henning
2000}. Here we describe a program of extensive observations of a new
sample of candidate HMPOs and present results indicating that some of
the sources studied are objects in the earliest evolutionary stage
known so far. Progress reports on this project were presented earlier
by
\citet{menten 1999}, \citet{sridharan 1999} and \citet{beuther 2000}.

\section {The Sample}
\label{sample_intro}

We have selected a sample of candidate HMPOs based on the criteria discussed
by \citet{rs}. They studied the reliability of the criteria defined by
\citet{wc89a} for identifying UCH{\sc ii} regions by their far-infrared 
properties and obtained a sample of relatively isolated HMPO
candidates in the process. The objects are chosen from the {\it IRAS}
Point Source Catalog and

\begin{itemize}
\item are detected in the high density gas tracing CS $J=2\to1$ survey 
  of massive star forming regions \citep{bronfman 1996}. These sources
  satisfy the criteria used by Wood \&\ Churchwell (1989) to select
  UCH{\sc ii} regions by their FIR colors.
\item are bright at FIR wavelengths ($F_{60} > 90 {\rm Jy} \land
  F_{100} > 500 {\rm Jy})$,
\item are not detected in the Galaxy-wide (single dish)
  5 GHz 1987 Green Bank \citep{parkes} and Parkes-MIT-NRAO radio
  continuum surveys \citep{griffith 1994,wright 1994} at flux
  densities above 25~mJy,

\item are north of $-$20$^\circ$ declination.
\end{itemize}

The first two of these criteria are based on the expectation that
dense gas is present and HMPOs and UCH{\sc ii} regions should resemble
each other with regard to their dust and gas temperatures and
luminosities. The third criterion ensures that we are dealing with
{\em isolated} HMPO candidates in the sense that they do not have
UCH{\sc ii} regions or more evolved H{\sc ii} regions in their
vicinity (within arcminutes, i.e. a few parsecs at the typical
distances of a few kpc). Ramesh \& Sridharan (1997) have shown that
the surveys under consideration are sensitive enough to detect the
radio continuum emission from typical UCH{\sc ii} regions throughout
the Galaxy. The last criterium ensures the observability with the
telescopes we are using.

These criteria led to the identification of 69 candidate objects with
kinematic distances \citep{brand 1993} derived from the CS
velocities. For sources inside the solar circle, there are two
solutions for the kinematic distance. This ambiguity can in some cases
be resolved by other means (see below). The sample is listed in
Table~\ref{sample} with systemic velocities, kinematic distances, and
luminosities and temperatures derived from the HIRES database (see
\S \ref{hires}). \citet{molinari 1996} used a different set of
far-infrared selection criteria to identify possible HMPO
candidates. Their and our sample have a total of 15 sources in common.

Figure \ref{distribution} (a and b) present plots of the Galactic
distribution and scale height for the whole sample.  Sources without
distance ambiguity are those in the outer Galaxy and those at the
tangential points. Furthermore, we solved the distance ambiguity for a
few more sources by other approaches: for sources with scale heights
over 170~pc at the far distance [$>4$ times the average Galactic scale
height of UCH{\sc ii} regions, 37~pc \citep{bronfman 2000}] the near
distance was adopted (18151$-$1208, 18272$-$1217 and 18517+0437); we
chose this scaleheight, because 166~pc is the largest offset from the
Galactic plane for a source with known distance (05553+1631). For
19217+1651 the far distance was adopted, because the observed VLA-cm
fluxes are not explicable by an implied low mass star if the near
distance is chosen. For a few objects distances of associated sources
were found in the literature (see Table~\ref{sample}).

We note that our selection criteria will certainly miss interesting
classes of HMPOs. The first class we would miss are the HMPOs that are
so young and cold that their SEDs peak longward of $\sim 200~\mu$m and
are extremely weak at near infrared-wavelengths, making them
undetectable by {\it IRAS} at 25~$\mu$m and thus excluding them from
our sample. For example, a 10$^4$~M$_{\odot}$ core of mean density of
$5\times 10^5$~cm$^{-3}$ and a temperature of 25~K, emitting as a
blackbody at a distance of 5~kpc, produces a 25~$\mu$m flux density
that is two orders of magnitude below the 0.5~Jy detection limit of
{\it IRAS} at that wavelength. A prototype of such a cold, young
object might be NGC 6334/I(NORTH) \citep{gezari 1982,megeath
1999}. The other missing group consists of sources with strongly
associated, but not coincident, mid-infrared sources which increase
the {\it IRAS} 12$\mu$m flux and decrease the 25$\mu$m/12$\mu$m flux
ratio, thus excluding them from our sample (e.g. 23385+6053,
\citealt{molinari 1998a}).

In this paper, the first of a series describing our HMPO survey, we
give an overview of the overall characteristics of the sample,
covering the mid-infrared ({\it IRAS} and MSX) and mm and cm continuum
and spectral line observations. Spatial structures are discussed and
we compare bolometric luminosities, gas masses and cm luminosities of
our sample with those of the ultracompact H{\sc ii} regions. H$_2$O
and CH$_3$OH maser observations as signposts of high-mass star
formation are presented and we discuss spectral line observations of
NH$_3$, CO, CH$_3$CN, thermal CH$_3$OH, SiO and H$^{13}$CO$^+$.

A more detailed analysis and discussion of the various datasets and
follow-up observations on individual sources will be published in
forthcoming papers.

\section{Observations and results}

\subsection{Archival data}

\subsubsection{The {\it IRAS}-HIRES database}

\label{hires}
The Infrared Astronomical Satellite ({\it IRAS}) performed an unbiased,
sensitive all sky survey at 12, 25, 60 and 100~$\mu$m. Advanced
processing of the survey data using the Maximum Correlation Method
resulted in HIRES (HIgh RESolution Processing) images, which have a
higher spatial resolution than the original {\it IRAS} dataset
\citep{aumann 1990}. We obtained images for the whole sample from 
the Infrared Processing and Analysis Center (IPAC) at 12, 25, 60, and
100~$\mu$m using 20 iterations of the algorithm. Spatial resolutions
are between $\approx 40''$ (12~$\mu$m) and 100$''$ (100~$\mu$m). These
are approximate values because the HIRES resolution varies around the
sky. Detailed descriptions of HIRES can be found on the IPAC
web-site\footnote{http://www.ipac.caltech.edu/ipac/iras/iras.html}.

In a majority of fields the 12~$\mu$m peak positions correspond well
with the 100~$\mu$m peaks within the positional uncertainties (for an
example see Fig.~\ref{hires2} bottom), but in 6 of the sources the
12~$\mu$m peaks are clearly offset by more than $1'$ from the
100~$\mu$m peaks (see Table \ref{results} and example in
Fig.~\ref{hires2} top). In some cases the 12~$\mu$m images are tracing
more evolved IR sources in the vicinity of our targets (e.g.
05358+3543, \citealt{porras 2000}). In other cases the 12~$\mu$m peaks
could be due to external heating of small grains (see discussion
below). A number of sources are resolved at the higher resolution of
the 12~$\mu$m images but appear unresolved at 100~$\mu$m
(e.g. 05553+1631, see Fig.~\ref{hires2} center). Thus, the fluxes in
the different bands in the {\it IRAS} point source catalog do not
necessarily correspond to the same objects. The real fraction of
non-coinciding 100 and 12~$\mu$m sources might be even higher, because
the spatial resolution, even of the HIRES images, is not very good.  To
improve on the Point Source Catalog data we obtained the fluxes around
the main dust cores
\citep{beuther 2001} from the HIRES images and fitted two-component
greybodies to these data (opacity corrected, modified blackbodies to
fit also the 1.2~mm data, \citealt{beuther 2001}). A cold dust
component fits the 60~$\mu$m and 100~$\mu$m data and a hot component
fits the 12~$\mu$m and 25~$\mu$m data. These fits reproduce the HIRES
data well (see Figure
\ref{fits_hires} for examples) for temperatures of the cold dust component
$\rm{T_{cd}}$ around 50~K and hot component $\rm{T_{hd}}$ between
150~K and 200~K (Table \ref{sample}).  While the hot component might
represent a more evolved and hot inner core, it could also be due to
small non-thermally excited grains exposed to external UV heating as
found for many other {\it IRAS} sources \citep{mathis 1990}. The MSX
data (see \S \ref{msx}) support this interpretation for a fraction of
sources for which the MSX- and mm positions do not
coincide. Integrating the two-component fits gives estimates of the
total luminosities of our objects (Table \ref{sample}). We find total
luminosities in the range $ 10^{3.5} - 10^6 L_{\odot}$, therefore we
conclude that the embedded sources are high-mass objects with ZAMS
masses $\gsim 8M_{\odot}$ (spectral type earlier than B2). The ratio
of the flux in the hot component to that in the cold component varies
between 0.05 and 0.4 with an average value around 0.16. This shows
that the main contribution to the total luminosity comes from the cold
components.

\subsubsection{The Midcourse Space Experiment (MSX) Point Source Catalog}

MSX, a Ballistic Missile Defense Organization satellite, surveyed the
whole Galactic plane (b$\pm 6^{\circ}$) in 6 bands centered at 4.29,
4.35, 8.28, 12.13, 14.65, and 21.34~$\mu$m. The sensitivity in the
4~$\mu$m bands is between 10 and 30~Jy/beam, in the other bands the
sensitivity is better, ranging between 0.1 and a few Jy/beam. With a
spatial resolution of 18.3$''$, positions in the Point Source Catalog
are accurate to an rms of $4''-5''$, and the calibration errors are
well within $13\%$.  For a more detailed description see, e.g.,
\citet{egan 1998} and the MSX web
-site\footnote{http://www.ipac.caltech.edu/ipac/msx/msx.html}.

The MSX Point Source Catalog lists sources in 62 out of our 69 fields
(Table \ref{sample}). In a few cases multiple sources are found and a
total of 87 sources are found toward our target regions.  Except for
18540+0220 and 23151+5912 no sources were detected in the 4~$\mu$m
bands, while in the other bands almost all of the sources are
detected. Comparing the fluxes in the 12~$\mu$m and 21~$\mu$m bands
(nearest to the 12$\mu$m \& 25$\mu$m {\it IRAS} bands) the flux ratio
is between 0.1 and 0.5 in 63 MSX sources, between 0.6 and 0.8 in 4
sources and undefined in 20 because of non-detection in either of the
two bands. The flux ratios between 0.1 and 0.5 are similar to the {\it
IRAS} flux ratios. For the main cores the flux ratios do not vary
between sources with or without cm detections, therefore the MSX
fluxes alone do not allow determinations of the evolutionary phase of
a source. The four sources with ratios in excess of 0.5 are found at
the edge or offset from the main core as seen in the mm continuum and
are most likely hotter and more evolved sub-sources.

\subsection{New observations}

\subsubsection{1.2~mm dust continuum emission}

Dust emission produced by massive cores has been found and mapped in
all the fields and a detailed analysis of this extensive database is
presented in an accompanying paper \citep{beuther 2001}. For the
present discussion we only use the core masses derived from the 1.2~mm data.

\subsubsection{3.6~cm continuum emission}

Because small H{\sc ii} regions are in general difficult to detect in
most existing centimeter-wavelength single dish surveys of the
Galactic plane, we used the Very Large Array (VLA)\footnote{The VLA is
operated by the National Radio Astronomy Observatory (NRAO) with
Associated Universities, Inc., under cooperative agreement with the
National Science Foundation (NSF).} to search for weak free-free
emission at 3.6~cm using the B array with a synthesized beam of
$0.7''$ and a $1\sigma$ rms sensitivity of $\sim$ 0.1~mJy. The observations
were carried out on July 2, 1998. Each source was observed twice at
widely different hour angles, for a total of 10 minutes in order to
improve $uv$-coverage. Suitable phase calibrators were observed at
intervals of approximately 30 minutes. The data were reduced using
NRAO's AIPS software using standard procedures.

Radio emission from objects in the pre-UCH{\sc ii} phase might be
caused by recently ignited massive protostars and/or jet phenomena
(e.g. \citealt{hofner 1999}). Background contamination by
extragalactic radio sources at a 1~mJy level is estimated to be
extremely low at this frequencies (0.007 per arcmin$^2$,
\citealt{fomalont 1991}). Other
processes such as synchrotron emission \citep{reid 1995} or stellar
winds could be producing the observed emission which require 
spectral index information for further elucidation.  About $10\%$ of the
sources originally observed by Wood \& Churchwell (1989b) are
optically thick at 2~cm, and a similar or even higher percentage
--~since in the early stages we expect to find hypercompact sources
($\leq 2000$~AU,
\citealt{tieftrunk 1997})~-- might be applicable for our
detections. Sources without any cm emission could be even younger
objects, where either no star has ignited or the ionized region is so
young and small that the free-free emission is still confined and not
detectable down to the mJy level \citep{churchwell 2000}. The mm
observations \citep{beuther 2001} prove that our sample indeed
consists of massive cores which should produce high-mass stars,
assuming normal IMFs. Out of the 66 fields observed at 3.6~cm, 24 show
no emission at the 1~mJy level, and 19 have weak emission with flux
densities below 10 mJy, while 21 sources show emission between 10 and
a few 100~mJy. For the remaining 2 fields the dynamic range of our
snapshot data is not good enough to draw a firm conclusion. Observed
flux densities are listed in Table
\ref{results}.  Typical ultracompact H{\sc ii} region fluxes at
distances of a few kpc are between a few mJy and more than a Jy
\citep{wc89b}, which is far higher than the flux densities observed in
a substantial fraction of our sample. But there are also a number of
well known sources which exhibit cm continuum emission at a level not
detectable with our sensitivity.  The W3(OH) Turner-Welch object (also
known as W3(H$_2$O)) at a distance of 2.2~kpc has a cm flux of 1.1~mJy
\citep{reid 1995} due to synchrotron emission, which is similar to our
detection limit, but other sources like source I in Orion KL
\citep{menten 1995} or Cep A East \citep{garay 1996} would show fluxes
below 0.3~mJy if moved to a distance of 2~kpc, which is below our
detection limit.  Therefore it is likely that deeper images of our
sample would reveal more cm sources. The fact that the sample contains
sources at different stages of evolution makes it an ideal
laboratory to explore the evolution of massive stars with a
consistent database.

\subsubsection{Spectral line observations} 

In order to understand the chemistry, kinematics and energetics of our
target regions, we conducted a number of different spectral line
observations from mm to cm wavelengths.

\subsubsubsection{Millimeter observations: outflow and dense core tracers}

The IRAM 30-m telescope was used (mainly in spring 1998, and for
several periods of a few hours duration in spring 1999 and 2000) to
obtain single point spectra for 36 sources in a number of tracers
sensitive to the physical and chemical conditions expected near HMPOs:
high temperatures and densities, elevated abundances of molecular
species resulting from evaporation of icy grain mantles, and outflow
activity. In most cases a number of lines could be observed
simultaneously (see Table \ref{line}).

The observations were conducted in the wobbler switching mode, where
the wobbling secondary reflector switched in azimuth between the ON
and an OFF position with a switching amplitude of $180''$ and a
frequency of 0.5~Hz. For the CO data the OFF position is generally not
free of emission, which corrupts the spectra at velocities around the
systemic velocity but in general does not affect spatially more
confined broad line wings in which we are primarily interested.  As
backends, we used the facility autocorrelator and two 512~MHz filterbanks.

To complete the CO line wing/outflow single positions search we
observed the rest of the sample in June 2000 at the CSO in the
$^{12}$CO$(2\to1)$ line (see Table \ref{line}). These data were
obtained in the position switching mode, where the OFF positions were
determined using Galactic CO surveys \citep{sanders 1986,dame 1987}
and verified to be free of emission.

We found CO line wings tracing bipolar outflows in 58 out of 69
sources. Six sources showed no line wings, and for 5 sources the
spectra are too confused by Galactic plane emission to allow a
classification. Assuming that line wings are only detectable for
outflow sources with an inclination angle of at least $10^{\circ}$ to
the plane of the sky, statistically $11\%$ (in our case $\approx 8$
sources) should show no wing emission although an outflow could be
present. That would be consistent with outflows being present in all
of our sources and strongly suggests that outflow activity is an
ubiquitous phenomenon not only in low mass star-forming regions but
also in regions of massive star formation. Previous investigations
that obtained similar results are e.g. \citet{shepherd 1996,henning
2000,zhang 2001} (see also the review by \citealt{churchwell 2000}).

SiO emission as another outflow tracer (e.g. \citealt{schilke 1997})
was detected in the 30-m observations toward 23 out of 36 sources with
wing emission seen in most cases.

Thermal emission from CH$_3$OH and/or CH$_3$CN was detected in 19 out
of the 30 sources implying that hot core activity is
prevalent. H$^{13}$CO$^+(1\to0)$ emission was found in all 36 observed
sources, most likely tracing structures similar to the dust
cores. We mostly used {\it IRAS} coordinates as source locations because at
the time of the observations improved positions from 1.2~mm
observations were not yet available. In 8 fields the main mm peak is
offset from the {\it IRAS} position by more than the 1.2~mm FWHM beamwidth
of $11''$, which would have lowered the detection rates.

\subsubsubsection{H$_2$O and CH$_3$OH masers}

During 1998 February and November, we searched for maser emission in
the 22~GHz H$_2$O (mean rms $\approx 0.4$~Jy) and 6.7~GHz CH$_3$OH
lines (mean rms $\approx 0.2$~Jy) toward all of our target sources
using the Effelsberg 100~m telescope (see Table \ref{line}). The
calibration error is estimated to be around $20\%$.

Both maser searches had high detection rates: 29 H$_2$O (10 of them
new detections) and 26 CH$_3$OH masers (5 new detections) were found;
toward 19 sources both H$_2$O and CH$_3$OH masers were detected (see
Fig. \ref{maserexamples} for examples). The CH$_3$OH detection rate
of our sample differs from that of UCH{\sc ii} regions: We find CH$_3$OH
maser toward $38\%$ of the sources in our sample while \citet{walsh
1998} quote a detection rate of $20\%$ toward UCH{\sc ii} regions. In
contrast, the detection rate of the H$_2$O masers toward
the sources of our sample is only 42$\%$, while \citet{churchwell 1990}
detect H$_2$O masers toward $67\%$ of 84 UCH{\sc ii} regions they
surveyed with comparable flux limits.

In Figure \ref{maserwc}, we present the source and maser distribution
of our sample in a Wood \& Churchwell (1989)-style color-color
diagram. While H$_2$O and CH$_3$OH maser sources are found all over
the diagram, combined clustering of both maser types is much more
common in the top-right corner, corresponding to the coldest and hence
presumably the least evolved sources, which makes sources with colors in that area
the most promising HMPO candidates.

While methanol maser emission is always confined to less than
15~km~s$^{-1}$ in total around the systemic velocity the water masers
nearly always emit over a much wider range (up to
70~km~s$^{-1}$). Just comparing the same velocity ranges both types
mostly show emission features at different velocities, which strongly
confirms earlier assumptions that these maser types are emitted in
different regions of the star-forming core \citep{menten 1996}. While
the methanol masers are probably confined to the innermost part --~the
molecular envelope of a central star or possibly a disk~-- water maser
emission could originate there as well, but the higher velocities point to 
different emission spots, most likely in outflows
(e.g. \citealt{marti 1999}).

Follow-up interferometric observations with the VLA (H$_2$O) and the
Australian Telescope Compact Array (ATCA, CH$_3$OH) have been
conducted and a detailed analysis and comparison of the data with the
other datasets will be published soon.

\subsubsubsection{NH$_3$ observations}

To obtain estimates of the kinetic temperature, the 100-m telescope
was used to observe the NH$_3$$(J,K) =$ (1,1), (2,2) inversion lines
in the entire sample (see Table \ref{line}).  The (1,1) line was detected
toward 59 sources, and for 40 sources the (2,2) line was found to be
strong enough to allow a meaningful estimate of the gas
temperature. The rotation temperatures listed in Table \ref{sample}
were derived following \citet{ungerechts 1986}; in the relevant
temperature range around 20~K kinetic temperatures are only marginally
higher
\citep{danby 1988}. The average rotation temperature is 19~K with the
notable exception of 18089-1732 for which we derive
38~K. \citet{molinari 1996} found a similar mean value of
$T_{\rm{kin}}\sim 22~K$ in their sample of young massive objects, which
is identical to the average rotation temperature \citet{churchwell
1990} found toward UCH{\sc ii} regions. Comparing the NH$_3$ derived
temperatures with the dust temperatures (from HIRES, Table
\ref{sample}) it is evident that the derived dust temperatures are on
the average $\sim 20$~K higher than the rotation temperatures. This
difference reflects the fact that {\it IRAS} was not sensitive to
temperatures lower than 30~K and therefore detected emission from warm
dust inside the core. In contrast to that the (1,1) and (2,2) ammonia
lines trace the cooler and more extended envelope and therefore result
in lower temperatures. It is possible and likely that NH$_3$(4,4)
observations will reveal higher temperatures due to a warm inner core.

The average intrinsic NH$_3$ line widths we find in our sample is
2.0~km~s$^{-1}$, while the average value found toward a sample of
molecular cores associated with UCH{\sc ii} regions is 3.1~km~s$^{-1}$
\citep{churchwell 1990}. This indicates significantly less turbulence
in the presumably younger HMPOs possibly because in this young sources
there has been less input of mechanical energy yet, e.g. by outflows.

\section{Discussion}

\subsection{Source Multiplicity}
\label{msx}

Positional comparison between MSX-/mm sources \citep{beuther 2001} and
MSX-/cm sources (Table \ref{results}) reveal median separations of
approximately $7''$ (between $1''$ and $30''$) in both cases. The
positional accuracy of MSX is $4''-5''$ \citep{egan 1998}, comparable
to that of the mm data, while the the cm data are more accurate, at better
than
$1''$. While the offsets mentioned might be due to positional
uncertainties in many cases, there remain significant numbers of
fields where the offsets are larger than any observational error and
therefore the different tracers depict different sources. MSX sources
located at the edges of the cores observed at mm-wavelengths might
either be reflection nebulae, i.e. reprocessed radiation from a deeply
embedded source, or they could also be sources belonging to the same
evolving cluster as the mm source, but in a more advanced stage of
evolution. Even younger stages of evolution might even be undetectable
at 12$\mu$m (see also
\S \ref{sample_intro}).

In 29 out of 42 fields with cm detections, MSX-, mm- and cm-sources
are found. In most of these cases it is difficult to draw firm
conclusions as to whether the emission in the three wavelength regimes
arises from the same or from different sources, because the spatial
resolution is insufficient. For the regions showing extended cm
emission (of size larger than a few arcsecs), in 5 out of 7 sources
the MSX-source is found to be associated with the cm- rather than with
the mm-source. While we are dealing with low-number statistics,
different stages of evolution in the same cluster are the most
probable cause. The more evolved and therefore hotter sources exciting
an UCH{\sc ii} region are more likely to emit at $12~\mu$m and thus be
detectable by MSX than the colder and more deeply embedded mm sources.

Generally, sources showing cm emission are found to be associated with
the mm cores \citep{beuther 2001}, but the cm peak is mostly offset
from the mm peak (Table \ref{results}). The median offset between mm
and cm sources is $11''$ (between $2''$ and $40''$). Given the
positional accuracies quoted above, these offsets are real and not due
to pointing problems. Assuming the near distance for sources with
distance ambiguities the median projected separation between mm and cm
peaks is $\approx 0.1$~pc. Thus, while the sources most likely belong
to the same (proto)cluster, they are clearly different
sub-sources. The cm sources are probably slightly older and more
evolved, while the mm sources might represent the youngest objects or
object groups in the fields. For the few sources which seem to be
coincident possible projection effects have also to be considered.

\subsection{Mass-luminosity relations}
\label{30m}

To understand mass and luminosity evolution in the massive star
formation process, we compare the masses derived from the mm continuum
data \citep{beuther 2001} and luminosities from the HIRES-data of our
sample with other classes of objects. Since we selected our sample to
contain objects in a pre-UCH{\sc ii} region phase, natural reference
objects are UCH{\sc ii} regions. For this purpose we use UCH{\sc ii}
region data from \citet{hunter 1997} and \citet{hunter 2000} and
estimate masses from 350$\mu$m images using the same assumptions as
employed by us \citep{beuther 2001} for our 1.2~mm data. For a few
sources Hunter et al.\ also have 1.3~mm data, and the masses separately
derived using only the 1.3mm fluxes and the 350$\mu$m fluxes agree well with 
each other.

In Figure \ref{comparison}(a) we plot far-infrared luminosities versus
core masses for HMPOs and UCH{\sc ii} regions. The two groups appear
separated, with the UCH{\sc ii} regions having higher bolometric
luminosities than the HMPOs for the same core masses. We interpret
this as an evolutionary effect caused by the embedded cluster, which
becomes more luminous at FIR wavelengths while they evolve and
destroy the cores from which they were born. Another possible
explanation for this trend would be that different massive star
clusters differ in their initial mass function (IMF) and that sources
with UCH{\sc ii} regions form more massive stars than HMPOs.  We
think this is unlikely, because Massey (1998) finds for more than 20 OB
associations essentially a Salpeter IMF ($N\propto m^{-1.35}$) with
little variation. Therefore, we argue that we are indeed observing an
evolutionary effect and our sample is on the average younger than
UCH{\sc ii} regions.

Monte Carlo simulations to derive expected cluster luminosities over a
wide range of cluster masses were performed using the methods outlined
in \citet{walsh 2001}. As Initial Mass Functions we used a Salpeter
IMF ($N\propto m^{-1.35}$) for sources above 1~M$_{\odot}$
\citep{massey 1998} and a Scalo IMF ($N\propto m^{-0.83}$) for sources
below 1~M$_{\odot}$. The upper end of the IMF is still under
discussion and \citet{casassus 2000} propose a steepening there, but
we work with the distribution found by \citet{massey 1998}, which is
based on observations of more than 20 OB associations. A steeper IMF
would flatten the top end of the derived cluster luminosity
distribution slightly, but qualitatively the results stay similar. As
mass-to-luminosity relation $L\propto m^{\beta}$ we use $\beta=2.8$
for $m<1$, $\beta=4$ for $1<m<30$ and $\beta=2$ for $m>30$
\citep{massey 1998,schatzman 1993}. To scale the cluster masses to the
initial core masses we assumed a star formation efficiency of $30\%$
\citep{lada 1993}, and the results are shown in greyscale in Figure
\ref{comparison}(a). The shaded area indicates where $90\%$ of the
star clusters are expected to be found. The large spread in luminosity
with given mass can be explained by the strong power law dependence
between mass and luminosity. Many low mass stars do not raise the
luminosity of the the cluster much, but the formation of just 1
massive star increases the cluster luminosity over orders of
magnitude. The upper cutoff at luminosities higher than
$10^6$~L$_{\odot}$ occurs because of an artificial upper mass
truncation in the simulated IMF at 100~M$_{\odot}$. The main
conclusion we derive from these simulations is that many HMPOs are
underluminous compared to the expected cluster luminosities. The
UCH{\sc ii} region luminosities agree better, but the most massive
UCH{\sc ii} regions are found to be underluminous again. Since we
expect the same formation processes in both types of sources with the
same underlying IMFs (as stated by \citealt{massey 1998}), this
strengthens our conclusion that the HMPOs are in a younger stage of
evolution and have not yet reached their final cluster luminosity.
The Figure suggests further that objects evolve from HMPOs to UCH{\sc
ii} regions to clusters.

A further support of the youth of our sample is outlined in Figure
\ref{comparison}(b), which presents the distance independent M/L ratio
in a histogram. As already expected from Figure \ref{comparison}(a)
the ratio M/L is lower on the average for UCH{\sc ii}s (peak at 0.01)
than for HMPOs (peak at 0.05). This again can be explained with the
HMPOs being younger and therefore less luminous than the more evolved
UCH{\sc ii} regions. A higher M/L ratio could also indicate sample of
less massive stellar objects on an average, but this is not the case
in our sample because the core mass ranges of the HMPOs and the
UCH{\sc ii} regions are similar (see Fig. \ref{comparison}(a)).

\subsection{Bolometric- versus cm luminosities}

We calculate from the measured cm fluxes the expected stellar
luminosities of the central sources 
[as outlined in \citet[eqs.~1~\&~3]{kurtz 1994} assuming
the emission to be optically thin at 3.6~cm] 
and compare them to the IR luminosities obtained from the {\it IRAS}
data base. For optically thick emission, which we expect in a fair
number of our sources, the derived luminosities are lower
limits. Figure \ref{luminosities} presents the derived cm luminosities
plotted against the corresponding IR luminosities.  It is obvious that
the IR luminosities are on the average far higher than the cm
luminosities; similar results were found by \citet{testi 2000} and
\citet{walsh 2001} on different samples.  This most likely reflects
the fact that the cm data trace just one single already ignited star
while the IR data --~due to the large {\it IRAS} beam~-- are tracing a
whole cluster with many sources, some younger with no H-burning at
their center yet, and others too weak to ionize a significant UCH{\sc
ii} region, which by now is the standard scenario. This would be
consistent with massive stars forming only in clusters \citep{stahler
2000}. A possible caveat is that some of the young ultracompact or
even hypercompact H{\sc ii} regions could be confined strongly and be
optically thick which leads to an underestimate of the cm
luminosities. Data at other wavelengths are necessary to define the
spectral indices and by that the ratio of optically thick
sources. Additionally, accretion contributions to the luminosities
have to be taken into account (see \S
\ref{accretion}). All these phenomena explain the observations
qualitatively well, but separating the different contributions
quantitatively is not possible with the present database.

\subsection{Accretion luminosity}\label{accretion}

Luminosity due to accretion plays a significant role in 
the formation of massive stars. We assume a simplified relation for
the accretion luminosity \citep{wolfire 1987}:

$$L_{\rm{acc}}=\frac{GM_{\ast}\dot{M}}{R_{\ast}}$$

with G the gravitational constant, $M_{\ast}$ the mass of the
(proto)star, $\dot{M}$ the accretion rate and $R_{\ast}$ the stellar
radius. Main sources of uncertainty are the accretion rate and the
radius of the accreting star. Assuming an accretion rate of
$\dot{M}=10^{-3}$~M$_{\odot}$/yr (based on outflow observations of
this sample presented in a forthcoming paper by Beuther et al., in
prep.) the accretion luminosity for stars between 10 and
40~M$_{\odot}$ is around $ 6\times 10^4~$L$_{\odot}$. While the
accretion rate we choose for this estimate is an upper limit in
current theories of high-mass star formation (e.g. \citealt{bernasconi
1996,norberg 2000,osorio 1999}), the radius of the evolving protostar
is expected to be larger than the final main sequence radius we
employed
\citep{bernasconi 1996}. Thus, these accretion luminosities are upper
limits to true accretion luminosities. With this assumed accretion
rates it takes $3\times 10^4$~years to build a 30~M$_{\odot}$ star
which has a luminosity comparable to the accretion luminosity. Using
the increasing accretion rates by \citet{norberg 2000} equality of
stellar and accretion luminosities is reached even later, at times
around $10^6$ years.
\citet{osorio 1999} find for a number of well known hot cores that
accretion still produces around $90\%$ of the bolometric luminosity
after a few times $10^4$ years.  In the models of intermediate-mass
stars by \citet{palla 1993} the radiative luminosity matches the
accretion luminosity after around $10^5$~years. While the time scales
differ between the different models, in all the cases the accretion
luminosity plays a dominant role for a significant amount of time
during the massive cluster forming process.  Now, comparing the
accretion luminosities with the IR derived luminosities of our sample
(Table \ref{sample}) it seems very likely that in many objects a
significant part of the detected flux is due to accretion. However, at
this stage of the investigation we cannot determine what fractions of
the total luminosity are due to accretion and nuclear burning. High
accretion rates
\citep{norberg 2000} would also cause quenching of the UCH{\sc ii} cm
radiation in the initial stages of evolution \citep{walmsley
1995,osorio 1999}, consistent with our low average cm fluxes.

\section{Conclusions}

Our selection criteria --~basically CS detections of bright {\it IRAS}
point sources with colors similar to ultracompact H{\sc ii} regions
and the absence of significant free-free emission at 5~GHz~-- were
very successful in finding sources at the earliest known stages of
high-mass star formation. An analysis of the {\it IRAS}-HIRES and
MSX-data gives an overview of spatial structures on large scales and
the spectral energy distributions. We find bolometric luminosities of
$10^4-10^6~\rm{L}_{\odot}$ and dust temperatures around 40~K.

\begin{itemize}

\item VLA cm observations reveal that a large fraction of the sample does
not emit free-free emission down to 1~mJy, which suggests that the sources
are at very early evolutionary stages. The majority of the detected
cm peaks is spatially offset from the mm peaks which implies different
sources in different evolutionary stages being responsible for the
different emission features. 

\item Luminosities based on the cm emission are
on the average far lower than bolometric luminosities derived in the
FIR. The most likely explanation for this phenomenon is that the cm
emission traces just one massive star whereas the FIR-emission
reflects the whole cluster luminosity. It has to be taken into account
that the cm emission might still be optically thick in a number of
sources, which results in an underestimation of the intrinsic
luminosity.

\item HMPOs are found to have a higher mass to luminosity ratio than UCH{\sc
ii} regions. The most likely explanation is that the HMPOs are
significantly younger with more molecular gas left and less (stellar)
radiation reprocessed in the IR. Evolving further to the UCH{\sc ii}
region stage the sources become more luminous and start destroying
their surrounding gas cocoons.

\item Comparing estimated accretion luminosities with the derived bolometric
luminosites indicates that in many sources a large fraction of the
total luminosity could be due to accretion processes.

\item Outflow phenomena are ubiquitous and methanol and water maser
detection rates are high. Other molecular species such as SiO (outflow
tracer), H$^{13}$CO$^+$ (dense gas tracer) and hot core tracers like
CH$_3$OH and CH$_3$CN were found in a large fraction of the observed
sources.

\item Rotation temperatures from NH$_3$ (1,1) and (2,2) lines in 40 out of
69 sources are on the average 20~K, about 20~K lower than the HIRES-derived
dust temperatures. This phenomenon has been found frequently and comes
about because {\it IRAS} was not sensitive to temperatures below 30~K and
observed warmer dust further inside the core. On the contrary, NH$_3$
(1,1) and (2,2) is sensitive to the cooler and more extended envelope,
which results in lower temperatures.

\end{itemize}

We have carried out further single dish and interferometric studies of
promising objects from this sample. These results will be published in
forthcoming papers (Beuther et al.\, Wyrowski et al.\ in preparation).
Combined with efforts by other groups a reasonable sized sample of
HMPO candidates now exists, and in the near future a good
observational characterization of HMPOs can be achieved to confront
theories.

\acknowledgements
We are indebted to Andrew Walsh for the Monte Carlo simulations of the
cluster luminosities. We thank Frank Bertoldi, Malcolm Walmsley and
Floris v.d. Tak for many useful comments on earlier versions of this
paper. Additionally, we thank Eberhard Hansis who worked on the HIRES
data and resolved some distance ambiguities. We also like to thank
Frederique Motte, Bernd Weferling and again Frank Bertoldi, who
conducted the bolometer observations at Pico Veleta. Last but not
least, we thank the referee for additional improving comments. The
observational work was conducted at the Effelsberg 100m, the VLA, the
IRAM 30m and the CSO (suppported by the NSF grant AST 96-15025). It
has made extensive use of the IPAC and SIMBAD databases. H. Beuther
gets support by the {\it Deutsche Forschungsgemeinschaft, DFG} project
number SPP 471. F. Wyrowski is supported by the National Science
Foundation under Grant No. AST-9981289.

\begin{deluxetable}{lrrrrrrrrrrrrr}
\tablecaption{The sample\label{sample}}
\tabletypesize{\tiny}
\startdata
source & R.A. & Dec.& v$_{\rm{lsr}}$ & d$_{\rm{far}}$ & d$_{\rm{near}}$ & T$_{\rm{cd}}$ & T$_{\rm{hd}}$ & L$_{\rm{far}}$  & L$_{\rm{near}}$ & 12/100 & T$_{\rm{rot}}$(NH$_3$) & MSX\\ 
    & [J2000] & [J2000] & [km/s]      & [kpc]       & [kpc]           &  [K]            & [K]            & log(L$_{\odot}$)& log(L$_{\odot}$) &[''] & [K] & \\
\tableline
 05358$+$3543  &  05 39 10.4  &  $+$35 45 19  &   $-$17.6  &  1.8$^1$  &        &  47  &  100  &  3.8  &     & 60          & 18  & $-$\\ 
 05490$+$2658  &  05 52 12.9  &  $+$26 59 33  &   0.8      &  2.1$^1$  &        &  44  &  167  &  3.5  &     &             &     & 1\\ 
 05553$+$1631  &  05 58 13.9  &  $+$16 32 00  &   5.7      &  2.5$^1$  &        &  56  &  114  &  3.8  &     &             &     & 1\\ 
 18089$-$1732  &  18 11 51.3  &  $-$17 31 29  &   33.8    &   13.0  &   3.6  &  40  &  113  &  5.6  &  4.5   & 100         & 38  &   4\\ 
 18090$-$1832  &  18 12 01.9  &  $-$18 31 56  &   109.8   &   10.0  &   6.6  &  36  &  157  &  4.5  &  4.1   &             &     & 1\\ 
 18102$-$1800  &  18 13 12.2  &  $-$17 59 35  &   21.1    &   14.0  &   2.6  &  35  &  161  &  5.3  &  3.8   &             & 15  & 1\\ 
 18151$-$1208  &  18 17 57.1  &  $-$12 07 22  &   32.8    &    3.0  &        &  47  &  170  &  4.3  &        &             & 17  & 1\\ 
 18159$-$1550  &  18 18 47.3  &  $-$15 48 58  &   59.9    &   11.7  &   4.7  &  55  &  123  &  5.0  &  4.2  &              &     & 1\\ 
 18182$-$1433  &  18 21 07.9  &  $-$14 31 53  &   59.1    &   11.8  &   4.5  &  43  &  118  &  5.1  &  4.3   &             & 20  & 1\\ 
 18223$-$1243  &  18 25 10.9  &  $-$12 42 17  &   45.5    &   12.4  &   3.7  &  50  &  173  &  5.3  &  4.2   &             & 18  & 1\\ 
 18247$-$1147  &  18 27 31.1  &  $-$11 45 56  &   121.7   &    9.3  &   6.7  &  35  &  132  &  5.0  &  4.8   &             &     & 2\\ 
 18264$-$1152  &  18 29 14.3  &  $-$11 50 26  &   43.6    &   12.5  &   3.5  &  35  &  151  &  5.1  &  4.0   &             & 18  & 1\\ 
 18272$-$1217  &  18 30 02.7  &  $-$12 15 27  &   34.0    &   2.9   &        &  52  &  181  &  4.0  &        &             &     & 1\\ 
 18290$-$0924  &  18 31 44.8  &  $-$09 22 09  &   84.3    &   10.5  &   5.3  &  40  &  165  &  5.0  &  4.4   &             & 20  & 1\\ 
 18306$-$0835  &  18 33 21.8  &  $-$08 33 38  &   76.8    &   10.7  &   4.9  &  34  &  154  &  4.8  &  4.1   &             &     & 1\\ 
 18308$-$0841  &  18 33 31.9  &  $-$08 39 17  &   77.1    &   10.7  &   4.9  &  38  &  142  &  4.9  &  4.2   &             & 18  & 1\\ 
 18310$-$0825  &  18 33 47.2  &  $-$08 23 35  &   84.4    &   10.4  &   5.2  &  40  &  129  &  4.8  &  4.2   &             & 18  & 3\\ 
 18337$-$0743  &  18 36 29.0  &  $-$07 40 33  &   57.9    &   11.5  &   4.0  &  37  &  139  &  5.0  &  4.0   &             & 17  & 1\\ 
 18345$-$0641  &  18 37 16.8  &  $-$06 38 32  &   95.9    & 9.5$^5$ &        &  36  &  161  &  4.6  &        &             & 16  & 1\\ 
 18348$-$0616  &  18 37 29.0  &  $-$06 14 15  &   109.5   &    9.0  &   6.3  &  39  &  159  &  5.1  &  4.8   &             & 19  & 3\\ 
 18372$-$0541  &  18 39 56.0  &  $-$05 38 49  &   23.6    &   13.4  &   1.8  &  33  &  204  &  5.3  &  3.5   & 110         & 16  & 1\\ 
 18385$-$0512  &  18 41 12.0  &  $-$05 09 06  &   26.0    &   13.1  &   2.0  &  49  &  157  &  5.3  &  3.7   &             &     & 1\\ 
 18426$-$0204  &  18 45 12.8  &  $-$02 01 12  &   15.0    &   13.5  &   1.1  &  32  &  204  &  5.0  &  2.8   &             &     & 1\\ 
 18431$-$0312  &  18 45 46.9  &  $-$03 09 24  &   105.2   &    8.2  &   6.7  &  34  &  199  &  4.5  &  4.4   &             & 15  & 2\\ 
 18437$-$0216  &  18 46 22.7  &  $-$02 13 24  &   110.8   &    7.3  &        &  31  &  221  &  4.4  &        &             &     & 2\\ 
 18440$-$0148  &  18 46 36.3  &  $-$01 45 23  &   97.6    &  8.3$^6$&        &  97  &  121  &  4.7  &        & 75          & 23  & 1\\ 
 18445$-$0222  &  18 47 10.8  &  $-$02 19 06  &   86.8    &    9.4  &   5.3  &  45  &  159  &  5.2  &  4.7   &             & 21  & 1\\ 
 18447$-$0229  &  18 47 23.7  &  $-$02 25 55  &   102.6   &    8.2  &   6.6  &  36  &  172  &  4.6  &  4.4   &             & 15  & 1\\ 
 18454$-$0136  &  18 48 03.7  &  $-$01 33 23  &   38.9    &   11.9  &   2.7  &  31  &  142  &  4.8  &  3.5   &             & 22  & 1\\ 
 18454$-$0158  &  18 48 01.3  &  $-$01 54 49  &   52.6    &  5.6$^7$&        &  34  &  170  &  4.3  &        &             &     & 3\\ 
 18460$-$0307  &  18 48 39.2  &  $-$03 03 53  &   83.7    &    9.5  &   5.2  &  43  &  176  &  4.6  &  4.1   &             & 19  & 3\\ 
 18470$-$0044  &  18 49 36.7  &  $-$00 41 05  &   96.5    & 8.2$^7$ &        &  56  &  158  &  4.9  &        &             & 20  & 2\\ 
 18472$-$0022  &  18 49 50.7  &  $-$00 19 09  &   49.0    &   11.1  &   3.2  &  49  &  135  &  4.9  &  3.8   &             &     & 1\\ 
 18488$+$0000  &  18 51 24.8  &  $+$00 04 19  &   82.7    &    8.9  &   5.4  &  42  &  169  &  4.9  &  4.5   &             & 20  & 1\\ 
 18517$+$0437  &  18 54 13.8  &  $+$04 41 32  &   43.9    &    2.9  &        &  38  &  176  &  4.1  &        &             &     & 1\\ 
 18521$+$0134  &  18 54 40.8  &  $+$01 38 02  &   76.0     &    9.0  &   5.0  &  37  &  151  &  4.6  &  4.1  & 90          &     & 1\\ 
 18530$+$0215  &  18 55 34.2  &  $+$02 19 08  &   77.7     &    8.7  &   5.1  &  51  &  127  &  5.4  &  4.9  &             & 16  & 1 \\ 
 18540$+$0220  &  18 56 35.6  &  $+$02 24 54  &   49.6     &   10.6  &   3.3  &  38  &  159  &  4.9  &  3.9  &             &     & 2 \\ 
 18553$+$0414  &  18 57 52.9  &  $+$04 18 06  &   10.0     &   12.9  &   0.6  &  42  &  152  &  5.1  &  2.4  &             &     & $-$ \\ 
 18566$+$0408  &  18 59 09.9  &  $+$04 12 14  &   85.2     &    6.7  &        &  51  &  161  &  4.8  &       &             & 15  & 1 \\ 
 19012$+$0536  &  19 03 45.1  &  $+$05 40 40  &   65.8     &    8.6  &   4.6  &  43  &  139  &  4.7  &  4.2  &             & 21  & 1 \\ 
 19035$+$0641  &  19 06 01.1  &  $+$06 46 35  &   32.4     & 2.2$^8$ &        &  51  &  100  &  3.9  &       &             & 21  & 1 \\ 
 19074$+$0752  &  19 09 53.3  &  $+$07 57 22  &   54.8     &    8.9  &   3.7  &  45  &  144  &  4.8  &  4.0  &             & 16  & 1 \\ 
 19175$+$1357  &  19 19 49.1  &  $+$14 02 46  &   14.6     & 10.6$^9$&        &  36  &  165  &  4.8  &       &             &     & 3\\ 
 19217$+$1651  &  19 23 58.8  &  $+$16 57 37  &   3.5      &   10.5  &        &  38  &  124  &  4.9  &       &             & 25  & $-$\\ 
 19220$+$1432  &  19 24 19.7  &  $+$14 38 03  &   68.8     &    5.5  &        &  43  &  151  &  4.4  &       &             & 23  & 1\\ 
 19266$+$1745  &  19 28 54.0  &  $+$17 51 56  &   5.0      &   10.0  &   0.3  &  32  &  176  &  4.7  &  1.7  &             &     & 2\\ 
 19282$+$1814  &  19 30 28.1  &  $+$18 20 53  &   23.6     &    8.2  &   1.9  &  36  &  231  &  4.9  &  3.6  &             &     & 1\\ 
 19403$+$2258  &  19 42 27.2  &  $+$23 05 12  &   26.7     &    6.3  &   2.4  &  57  &  150  &  4.7  &  3.8  &             &     & 1\\ 
 19410$+$2336  &  19 43 11.4  &  $+$23 44 06  &   22.4     &    6.4  &   2.1  &  46  &  147  &  5.0  &  4.0  &             & 18  & 1\\ 
 19411$+$2306  &  19 43 18.1  &  $+$23 13 59  &   29.0     &    5.8  &   2.9  &  41  &  157  &  4.3  &  3.7  &             & 14  & 1\\ 
 19413$+$2332  &  19 43 28.9  &  $+$23 40 04  &   20.8     &    6.8  &   1.8  &  41  &  133  &  4.4  &  3.3  &             & 18  & 1\\ 
 19471$+$2641  &  19 49 09.9  &  $+$26 48 52  &   21.0     &2.4$^{10}$&       &  38  &  194  &  3.6  &       & 130         &     & 2\\ 
 20051$+$3435  &  20 07 03.8  &  $+$34 44 35  &   11.6     &    3.7  &   1.6  &  49  &  143  &  4.0  &  3.3  &             &     & 1\\ 
 20081$+$2720  &  20 10 11.5  &  $+$27 29 06  &   5.7      &  0.7$^2$  &        &  34  &  248  &  2.5  &     &             &     & 1\\ 
 20126$+$4104  &  20 14 26.0  &  $+$41 13 32  &   $-$3.8   &  1.7$^2$  &        &  62  &  112  &  3.9  &     &             & 23  & 1\\ 
 20205$+$3948  &  20 22 21.9  &  $+$39 58 05  &   $-$1.7   &    4.5  &        &  48  &  175  &  4.5  &       &             &     & 1\\ 
 20216$+$4107  &  20 23 23.8  &  $+$41 17 40  &   $-$2.0   &  1.7$^2$  &        &  46  &  159  &  3.3  &     &             & 21  & 1\\ 
 20293$+$3952  &  20 31 10.7  &  $+$40 03 10  &   6.3      &    2.0  &   1.3  &  56  &  157  &  3.8  &  3.4  &             & 15  & 1\\ 
 20319$+$3958  &  20 33 49.3  &  $+$40 08 45  &   8.8      &    1.6  &        &  73  &  162  &  3.8  &       &             &     & 1\\ 
 20332$+$4124  &  20 35 00.5  &  $+$41 34 48  &   $-$2.0   &    3.9  &        &  56  &  148  &  4.4  &       &             & 17  & $-$\\ 
 20343$+$4129  &  20 36 07.1  &  $+$41 40 01  &   11.5     &    1.4  &        &  44  &  150  &  3.5  &       &             & 18  & $-$\\ 
 22134$+$5834  &  22 15 09.1  &  $+$58 49 09  &   $-$18.3  &    2.6  &        &  61  &  135  &  4.1  &       &             & 18  & $-$\\ 
 22551$+$6221  &  22 57 05.2  &  $+$62 37 44  &   $-$13.4  &  0.7$^3$  &        &  45  &  186  &  3.2  &     &             &     & $-$\\ 
 22570$+$5912  &  22 59 06.5  &  $+$59 28 28  &   $-$46.7  &    5.1  &        &  54  &  145  &  4.7  &       &             &     & 1\\ 
 23033$+$5951  &  23 05 25.7  &  $+$60 08 08  &   $-$53.1  &  3.5$^4$  &        &  52  &  149  &  4.0  &     &             & 20  & 1\\ 
 23139$+$5939  &  23 16 09.3  &  $+$59 55 23  &   $-$44.7  &    4.8  &        &  41  &  140  &  4.4  &       &             &     & 1\\ 
 23151$+$5912  &  23 17 21.0  &  $+$59 28 49  &   $-$54.4  &    5.7  &        &  68  &  175  &  5.0  &       &             &     & 1\\ 
 23545$+$6508  &  23 57 05.2  &  $+$65 25 11  &   $-$18.4  &  0.8$^2$  &        &  50  &  181  &  3.0  &     &             & 18  & 1\\ 
\enddata\\
\tablecomments{Only far and no near distance means that the distance ambiguity is solved. Rotation temperatures T$_{\rm{rot}}$(NH$_3$) are correct to approximately 3~K. The 12/100 column refers to offsets between the corresponding IRAS sources and the last columns shows the number of MSX point source catalog detections in fields of radius $2'$ around the {\it IRAS} position. $^1$ Snell et al. 1990; $^2$ Wilking et al. 1989; $^3$ Blaauw 1964; $^4$ Jijina et al. 1999;$^5$ Garay et al. 1993; $^6$ Walsh et al. 1997; $^7$ Kuchar \& Bania 1994; $^8$ Osterloh et al. 1996; $^9$ near star HD231160; $^{10}$ near star CGO 544}
\end{deluxetable}

\begin{deluxetable}{lrrrrrr}
\tablecaption{Observing parameters\label{line}}
\tabletypesize{\scriptsize}
\startdata
\hline                         
& freq.  & HPBW & $T_{\rm{sys}}$ & $\Delta v$ & Obs. & tracer \\ &
[GHz] & [$''$] & [K] & [km~s$^{-1}$] & \\ $^{12}$CO $2\to1$ & 230.5 &
11 & 250 & 0.1 & PV & outflow\\ SiO $2\to1$ & 86.9 & 29 & 85 & 3 & PV
& outflow \\ H$^{13}$CO$^+$ $1\to0$ & 86.9 & 29 & 85 & 3 & PV & dense
core\\ $^{13}$CO $1\to0$ & 110.2 & 22 & 120& 0.8 & PV & dense core\\
CH$_3$OH $J_k=5_k\to4_k$ & $\sim 241.8$ & 11 & 250& 0.1 & PV & dense
core\\ CH$_3$CN $J=6\to5$ & $\sim 110.4$& 22 & 120 & 0.8 & PV & dense
core\\ $^{12}$CO $2\to1$ & 230.5 & 27 & 500 & 0.06& CSO & outflow\\
NH$_3$ (1,1), (2,2) & $\sim 22.6$ & 40 & 50 & 0.25& Eff.  & dense
core\\ H$_2$O maser & 22.2 & 40 & 50 & 0.03& Eff.  & outflow and/or
disk ?\\ CH$_3$OH maser & 6.7 & 130 & 50 & 0.1 &Eff.  & high-mass
signpost\\
\enddata
\tablecomments{Observatories: Eff.: Effelsberg, PV: Pico veleta, CSO: Caltech Submillimeter observatory}
\end{deluxetable}

\begin{deluxetable}{lrrrrrrrrrr}
\tablecaption{Results \label{results}}
\tabletypesize{\tiny}
\startdata
source & cm flux & H$_2$O.& CH$_3$OH                   & MSX/cm & MSX/mm & cm/mm & wings & SiO & CH$_3$OH & CH$_3$CN \\ 
       & [mJy]   & [Jy]   & [Jy]                       & $['']$ & $['']$ & $['']$&       &     &          &          \\ 
\tableline					      	  
 05358$+$3543  &  $<1$  &   45     &   162.0$^1$       &     &        &      & + & +  & +  & -- \\ 
 05490$+$2658  &  $<1$  &   $-$     &   $-$            &     &   16.9 &      & + & -- & -- & -- \\  
 05553$+$1631  &  1.3  &   $-$      &   $-$            & 3.5 &    0.8 & 2.8  & + & +  & +  & +  \\  
 18089$-$1732  &  0.9  &   75       &   54.0 $^2$      & 8.1 &   11.2 & 3.2  & + & +  & +  & +  \\  
 18090$-$1832  &    ?  &   $-$           &   82.0$^2$  &     &    2.6 &      & + &    &    &    \\ 
 18102$-$1800  &   44  &   $-$           &   8.8$^3$   & 1.7 &   25.3 & 27.0 & + & +  & -- & -- \\ 
 18151$-$1208  &  $<1$  &   0.8$^*$    &   50.0$^4$    &     &    8.3 &      & + & +  & +  & +  \\ 
 18159$-$1550  &  10.0  &   ?            &   $-$       & 4.4 &   16.9 & 12.6 & + &    &    &    \\ 
 18182$-$1433  &  $<1$  &   18         &   24.0$^2$    &     &   10.8 &      & + & +  & +  & +  \\  
 18223$-$1243  &  $<1$  &   $-$           &   $-$      &     &    3.6 &      & + & -- & -- & -- \\ 
 18247$-$1147  &   47  &   $-$           &   1.6       & 7.4 &    5.4 & 1.9  & + &    &    &    \\ 
 18264$-$1152  &  $<1$  &  50$^*$      &   3.8$^1$     &     &    9.4 &      & + & +  & +  & +  \\  
 18272$-$1217  &  110  &   $-$           &   $-$       & 4.4 &    8.9 & 5.3  & --&    &    &    \\ 
 18290$-$0924  &  7.0  &   4        &   15.1$^2$       &12.7 &    4.5 & 17.3 & + & -- & -- & -- \\ 
 18306$-$0835  &   82  &   0.7$^*$    &   $-$          &15.5 &   11.6 & 11.8 & + &    &    &    \\ 
 18308$-$0841  &  $<1$  &  1.5$^*$    &   $-$          &     &   10.6 &      & --& +  & -- & +  \\  
 18310$-$0825  &  7.0  &   $-$         &   $-$         & 5.1 &   16.8 & 10.6 & + &    &    &     \\ 
 18337$-$0743  &  $<1$  &   $-$           &   $-$      &     &    4.7 &      & ? &    &    &     \\ 
 18345$-$0641  &   27  &   3$^*$     &   5.4$^4$       &     &    4.0 &      & + & +  & +  & +   \\ 
 18348$-$0616  &   54  &   $-$         &   $-$         &14.8 &   18.6 & 67.5 & + & -- & -- & --  \\ 
 18372$-$0541  &   80  &   1.5$^*$      &   9.2$^1$    & 3.0 &    1.8 & 1.9  & + &    &    &     \\ 
 18385$-$0512  &   29  &   200         &   $-$         & 4.4 &    2.8 & 3.4  & + &    &    &   \\  
 18426$-$0204  &  1.6  &   ?          &   1.3          & 6.7 &   12.0 & 17.6 & + &    &    &   \\  
 18431$-$0312  &  3.5  &   $-$           &   $-$       &30.3 &    2.4 & 32.0 & + &    &    &    \\  
 18437$-$0216  &  $<1$  &   $-$         &   $-$        &     &    6.9 &      & ? &    &    &    \\ 
 18440$-$0148  &  $<1$  &    ?         &   4.0$^2$     &     &    2.6 &      & + & -- & -- & --  \\ 
 18445$-$0222  &   68  &   $-$           &   $-$       & 0.4 &   18.5 & 18.9 & ? &    &    &     \\ 
 18447$-$0229  &  1.6  &   $-$           &   $-$       &13.2 &    4.4 & 10.0 & ? &    &    &     \\ 
 18454$-$0136  &   42  &   $-$         &   2.5$^1$     &     &    4.6 & 6.7  & + &    &    &    \\ 
 18454$-$0158  &   18  &   $-$           &   $-$       &     &        & 2.2  & + &    &    &   \\  
 18460$-$0307  &  $<1$  &   $-$           &   $-$      &     &    5.3 &      & + &    &    &     \\ 
 18470$-$0044  &    ?  &   $-$         &   7.5$^3$     &     &   20.6 &      & + & +  & +  & +  \\  
 18472$-$0022  &  110  &   $-$         &   $-$         &     &    2.1 & 0.8  & + &    &    &    \\  
 18488$+$0000  &  194  &  1$^*$   &   16.9$^4$         &19.0 &    4.1 & 16.2 & + & +  & +  & +  \\ 
 18517$+$0437  &  $<1$  &   45.3     & 279$^1$          &     &   22.6 &      & + &    &    &    \\ 
 18521$+$0134  &  $<1$  &   $-$     &   1.3            &     &    3.7 &      & --&    &    &    \\  
 18530$+$0215  &  311  &   $-$     &   $-$             & 7.3 &    7.2 & 3.8  & + & +  & +  & +   \\ 
 18540$+$0220  &   97  &   $-$       &   $-$           &12.1 &    2.8 & 10.3 & ? &    &    &    \\ 
 18553$+$0414  &  $<1$  &  50$^*$   &   $-$            &     &        &      & + &    &    &    \\ 
 18566$+$0408  &  $<1$  &  3$^*$   &   7.2$^3$         &     &    0.9 &      & + & +  & +  & +   \\ 
 19012$+$0536  &  $<1$  &   2$^*$   &   0.9            &     &    5.9 &      & + & +  & +  & --  \\ 
 19035$+$0641  &  4.0  &   9       &   14.3$^5$        & 5.1 &    7.1 & 9.0  & + & +  & +  & --  \\ 
 19074$+$0752  &  14.8  &   $-$       &   $-$          &     &    2.3 &      & + &    &    &     \\ 
 19175$+$1357  &  5.1  &   $-$     &   $-$             &     &    2.1 & 37.0 & --&    &    &    \\ 
 19217$+$1651  &   32  &  9$^*$   &   0.9              &     &        & 6.5  & + & +  & +  & +   \\ 
 19220$+$1432  &  11.0  &   $-$       &   $-$          & 4.8 &    6.1 & 10.4 & + & -- & -- & --  \\ 
 19266$+$1745  &  $<1$  &  1.5$^*$  &   1.7$^1$        &     &    4.1 &      & + & -- & -- & --\\  
 19282$+$1814  &  $<1$  &   $-$     &   2.9$^1$        &     &   11.8 &      & + &    &    &   \\  
 19403$+$2258  &  $<1$  &   $-$       &   $-$          &     &   23.5 &      & + &    &    &   \\  
 19410$+$2336  &  1.0   &   110   &   30.0$^1$          &     &    3.8 &      & + & +  & +  & +  \\  
 19411$+$2306  &  $<1$  &   $-$       &   $-$          &     &    2.5 &      & + & +  & +  & --  \\ 
 19413$+$2332  &  $<1$  &   $-$       &   $-$          &     &   11.1 &      & + & +  & +  & -- \\ 
 19471$+$2641  &    ?  &   $-$     &   $-$             &     &        &      & + &    &    &   \\  
 20051$+$3435  &  $<1$  &   $-$     &   $-$            &     &    6.7 &      & --&    &    &     \\ 
 20081$+$2720  &  $<1$  &   $-$   &   $-$              &     &        &      & + &    &    &    \\ 
 20126$+$4104  &    ?  &   15     &   36.0$^6$         &     &    2.6 &      & + & +  & +  & +   \\ 
 20205$+$3948  &  8.3  &   $-$       &   $-$           &12.1 &   11.5 & 23.0 & + & -- & -- & -- \\ 
 20216$+$4107  &  1.4  &   $-$       &   $-$           &     &    4.3 &      & + & -- & -- & --  \\ 
 20293$+$3952  &  7.6  &  100$^*$  &   $-$             & 2.0 &   23.5 & 15.0 & + & +  & +  & +  \\  
 20319$+$3958  &   25  &   $-$     &   $-$             & 8.3 &    8.5 & 15.5 & + &    &    &    \\ 
 20332$+$4124  &  136  &  0.6$^*$  &   $-$             &     &        & 10.5 & + &    &    &     \\ 
 20343$+$4129  &  1.8  &    $-$  &   $-$               &     &        & 13.2 & + & -- & -- & --  \\ 
 22134$+$5834  &  3.7  &   $-$      &   $-$            &     &        & 3.5  & + & -- & -- & --  \\ 
 22551$+$6221  &  4.5  &   $-$    &   $-$              &     &        & 74.7 &-- &    &    &    \\ 
 22570$+$5912  &   29  &   $-$        &   $-$          & 6.4 &   13.3 & 7.3  & + & -- & -- & -- \\ 
 23033$+$5951  &  1.7  &   4        &   $-$            & 6.4 &   10.6 & 7.2  & + & +  & +  &--   \\ 
 23139$+$5939  &  1.4  &   400   &   2.6$^1$           & 2.1 &    1.2 & 1.0  & + & +  & +  & +  \\ 
 23151$+$5912  &  $<1$  &   60       &   $-$           &     &    7.7 &      & + & +  & -- & -- \\ 
 23545$+$6508  &  1.0  &   $-$   &   $-$               &13.0 &   26.3 & 24.2 & + & +  & -- & --  \\ 
\enddata\\							       
\tablecomments{'$^*$': new detections, '?': tentative detections; the MSX/cm MSX/mm and cm/mm columns present offsets between the different observations, and the last 4 column show detections (+), non-detection (--) or non-observations (empty space); $^1$ Szymczak et al. 2000; $^2$ Walsh et al. 1998; $^3$ Slysh et al. 1999; $^4$ v. d. Walt et al. 1995; $^5$ Caswell Vaile 1995; $^6$ MacLeod 1992}
\end{deluxetable}

\includegraphics[angle=-90,width=16cm]{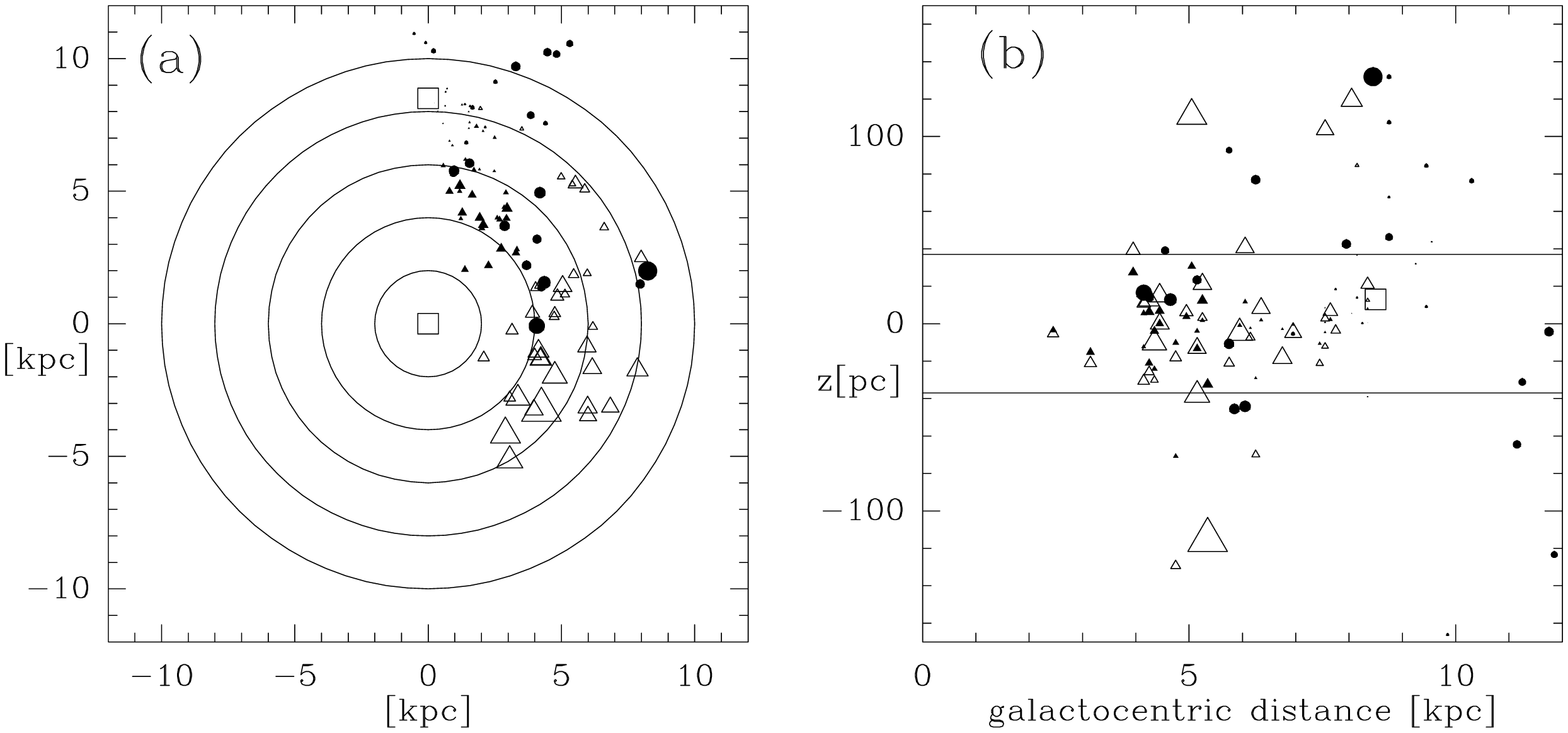}
\figcaption[spiral.ps]{{\bf (a)} The Galactic distribution of the sample: filled circles represent sources with no distance ambiguity, triangles show sources with distance ambiguities (filled triangles: near; open triangles: far). The size of the symbols scales with their luminosity. Squares show the Galactic center and the location of the sun (8.5~kpc from the center), while the circles indicate galactic distances in 2~kpc steps. {\bf (b)} The Galactic scale height distribution of the sample. Symbols are the same as in (a). The horizontal lines show the average Galactic scale height of massive star formation regions 37~pc (Bronfman et al.\ 2000). \label{distribution}}

\newpage
\includegraphics[angle=-90,width=11cm]{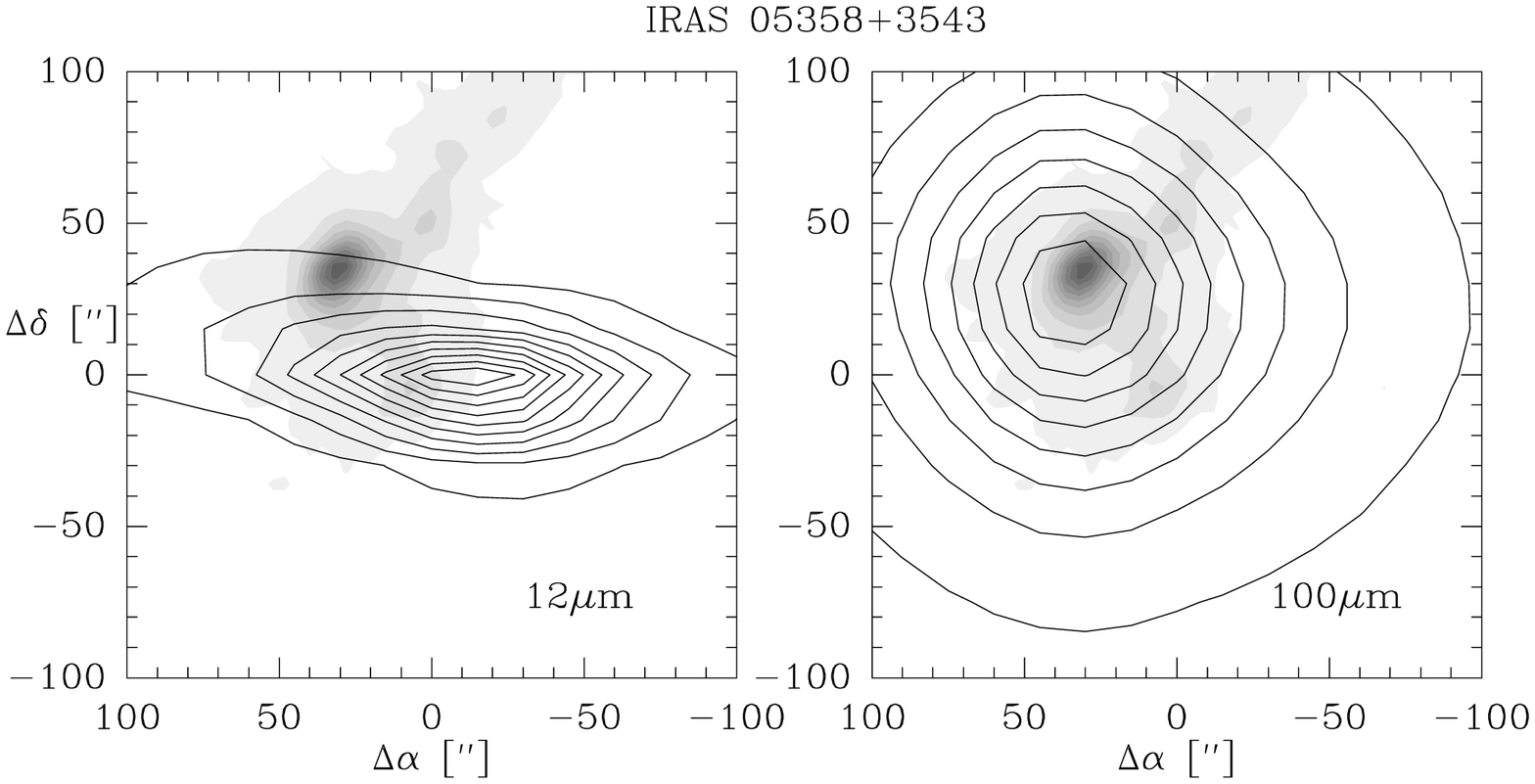}\\
\indent \includegraphics[angle=-90,width=11cm]{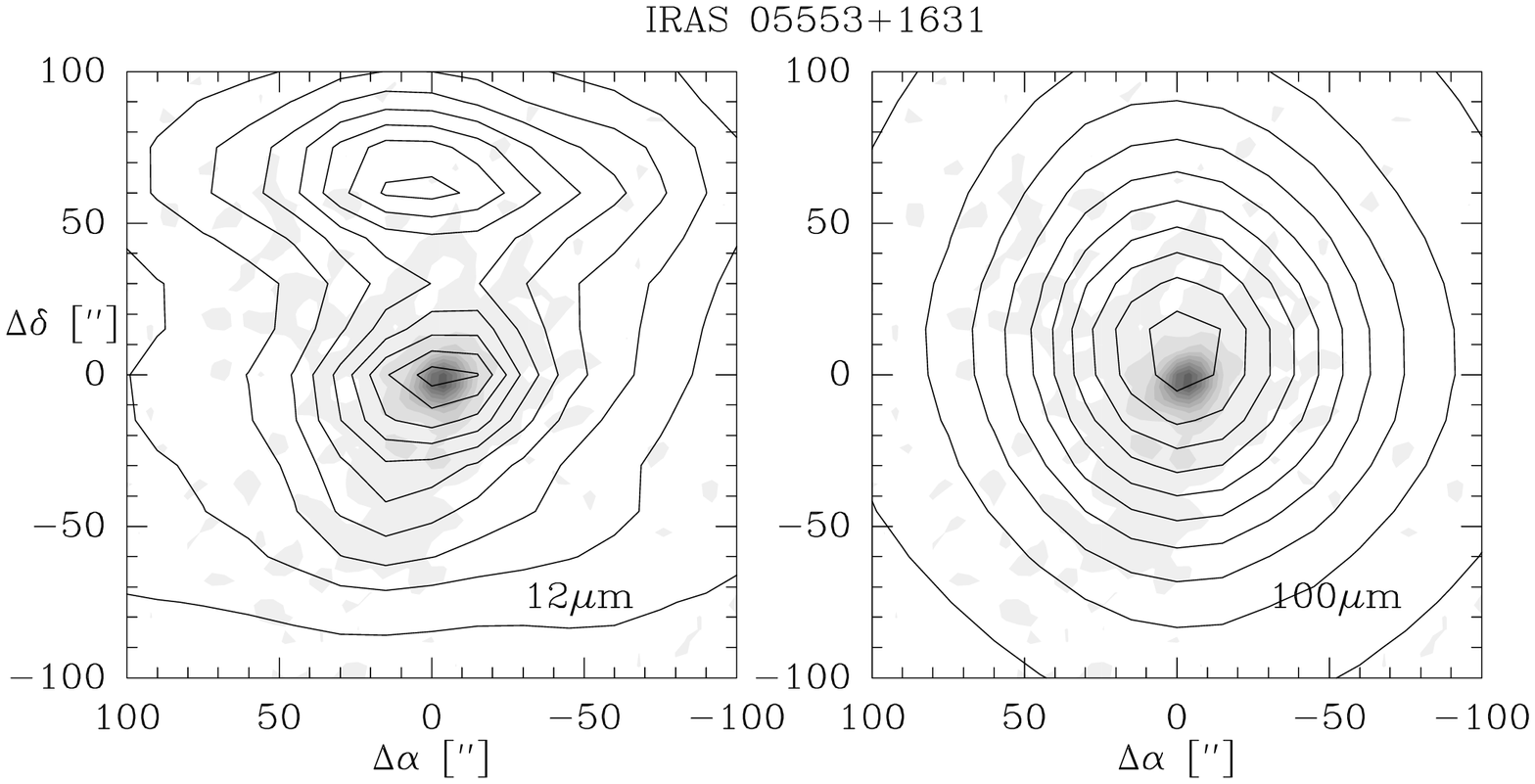}\\
\indent \includegraphics[angle=-90,width=11cm]{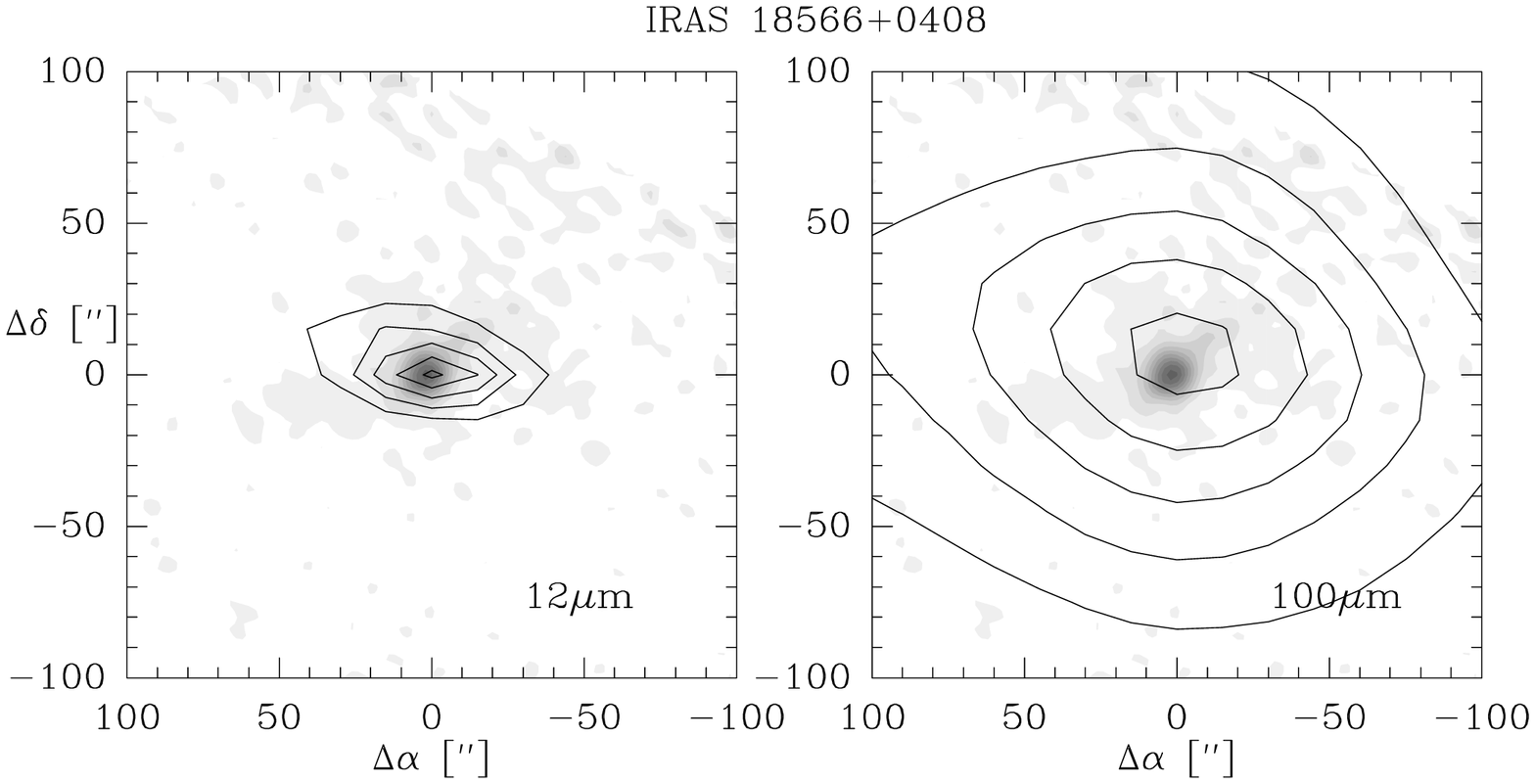}
\figcaption[]{Examples of the HIRES images (contours) at 12~$\mu m$ and 100~$\mu m$ on greyscale 1.2~mm dust images \citep{beuther 2001}. While the bottom images are typical with both HIRES bands corresponding to the same mm core, the middle and top images are counterexamples with the 12~$\mu m$ peaks splitting up or tracing different sources in the same field of view. \label{hires2}}

\newpage
\includegraphics[angle=-90,width=8cm]{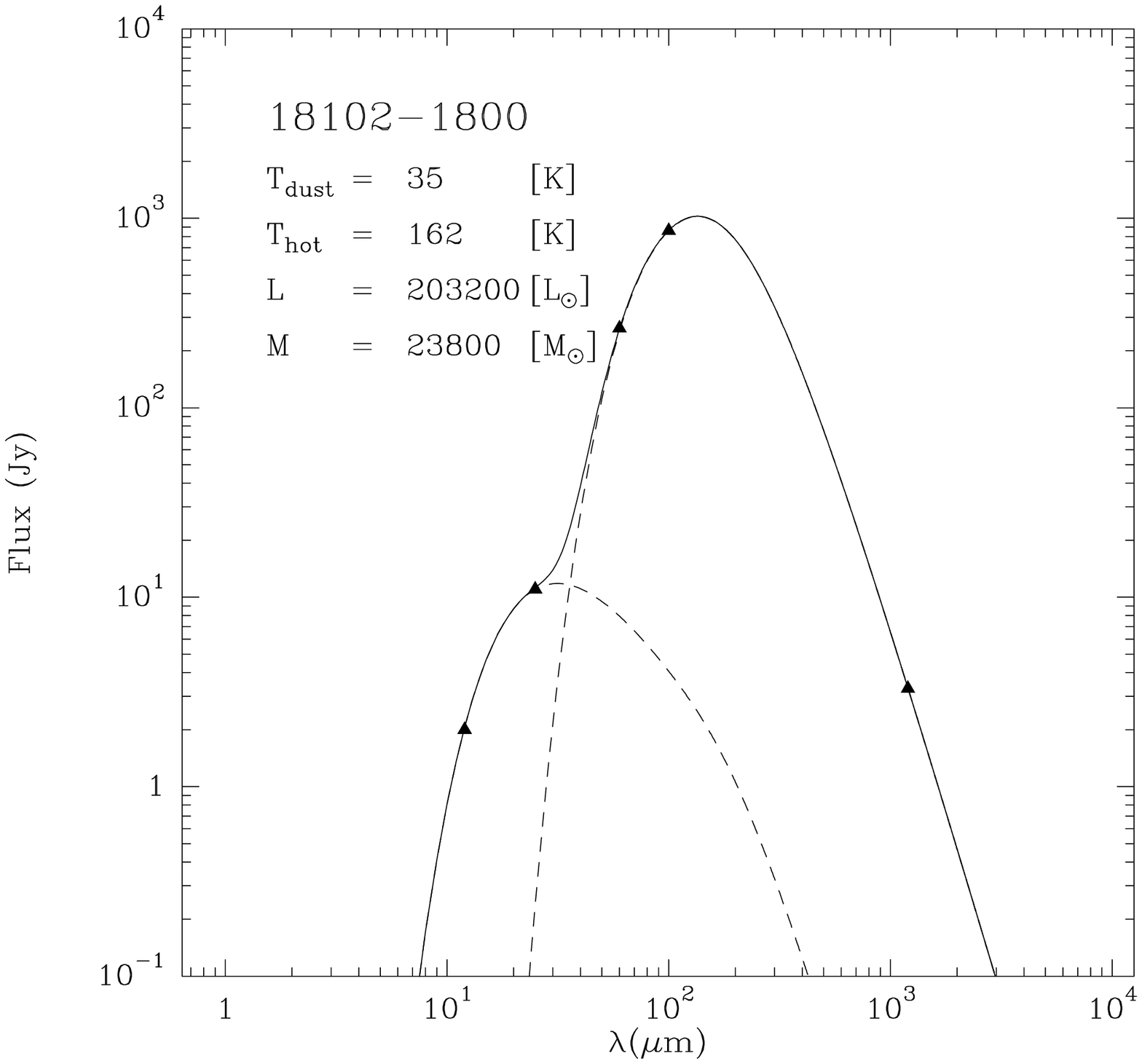}
\includegraphics[angle=-90,width=8cm]{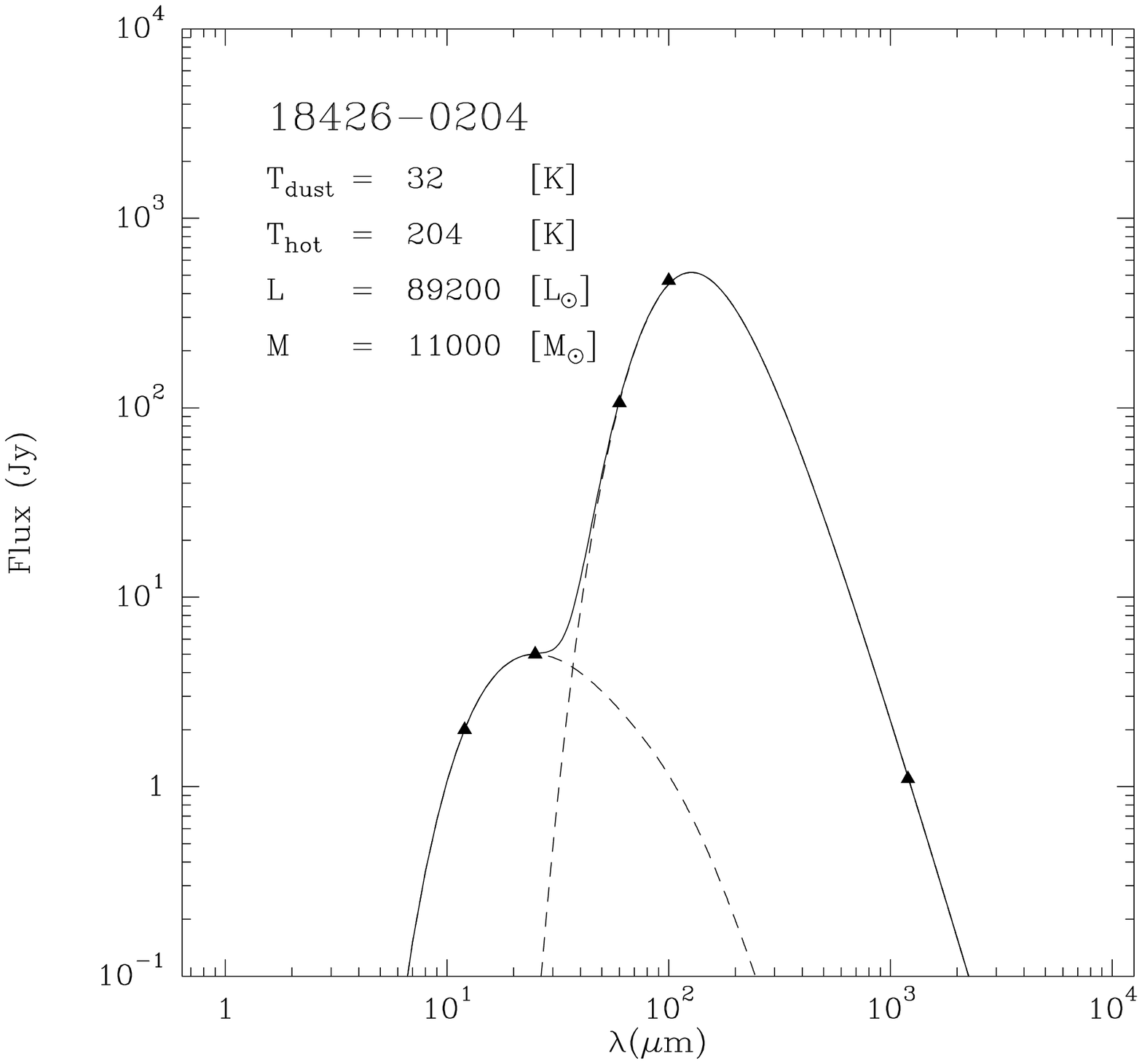}
  \figcaption[fits_hires.ps]{Examples of HIRES 2 component greybody
  fits. While the dashed lines show the two greybodies separately, the
  full line shows the two component fit. Each panel presents on the
  top left the source name, luminosities and
  temperatures. \label{fits_hires}}
 
\newpage
\includegraphics[angle=-90,width=7.8cm]{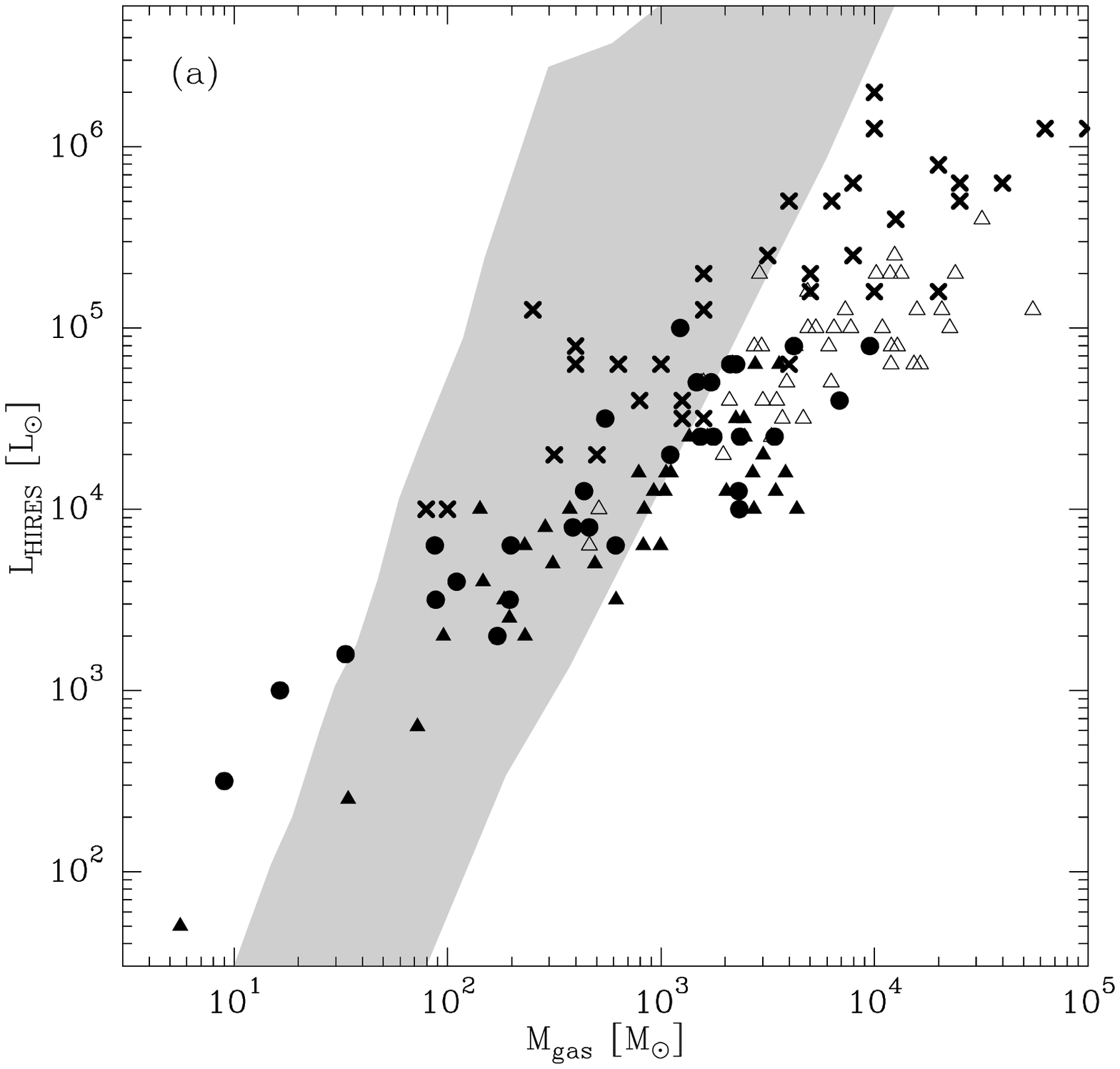}
\includegraphics[angle=-90,width=7.6cm]{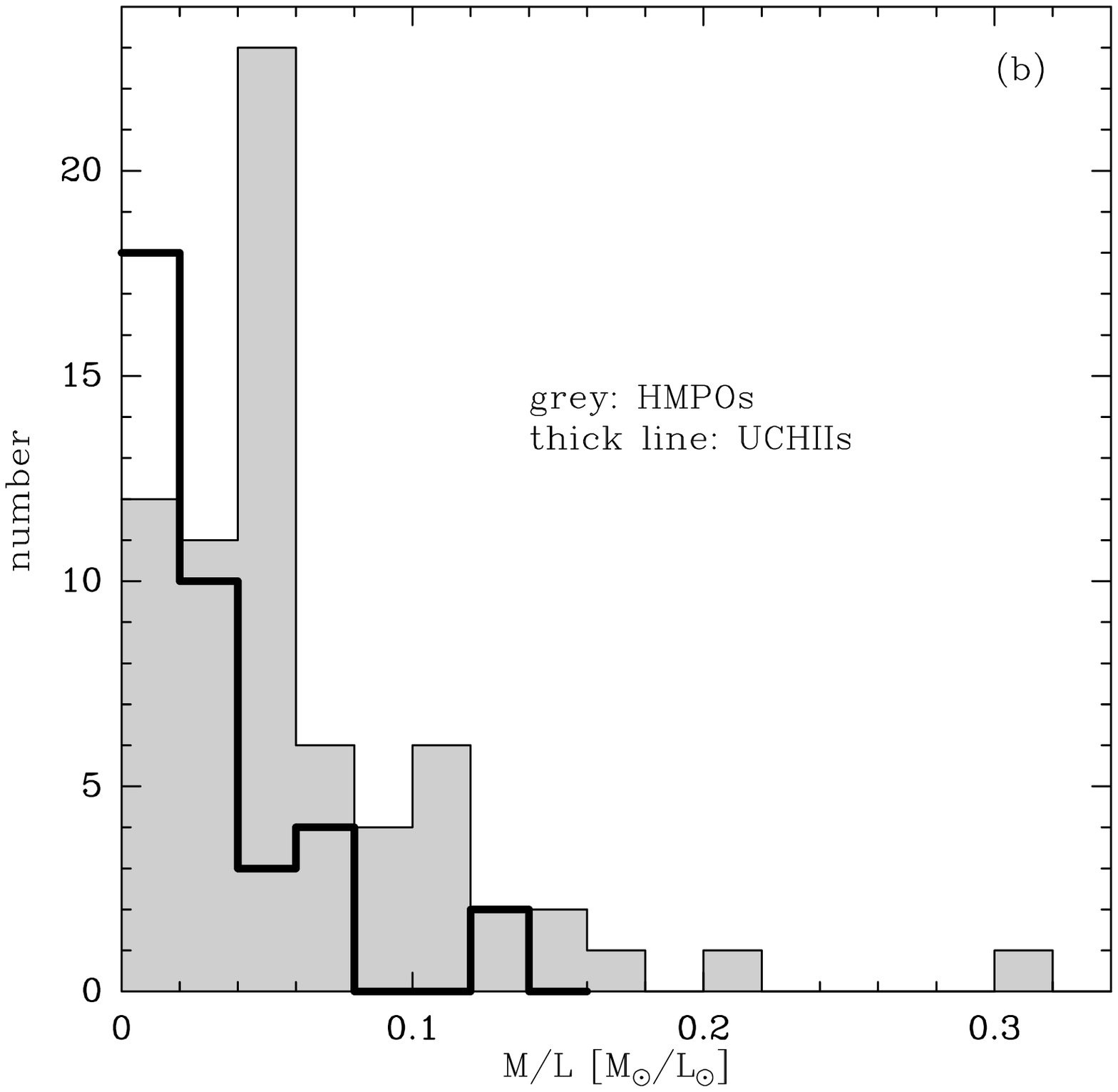}\\
  \figcaption[l_m.ps]{{\bf (a):} Plotted are the derived FIR
  luminosities against the core masses. Triangles show HMPOs with
  distance ambiguity (filled: near distance; open: far distance) and
  filled circles without distance ambiguity. The crosses show data
  from UCH{\sc ii} regions (Hunter et al. 1997, 2000). The grey shaded
  area defines the region in which $90\%$ of the simulated
  clusters are found \citep{walsh 2001}. {\bf (b):} The histogram
  presents the distance independent quantity M/L with again a clear
  separation of HMPOs (grey) and UCH{\sc ii}s (thick
  line). \label{comparison}}

\newpage 
\includegraphics[angle=-90,width=16cm]{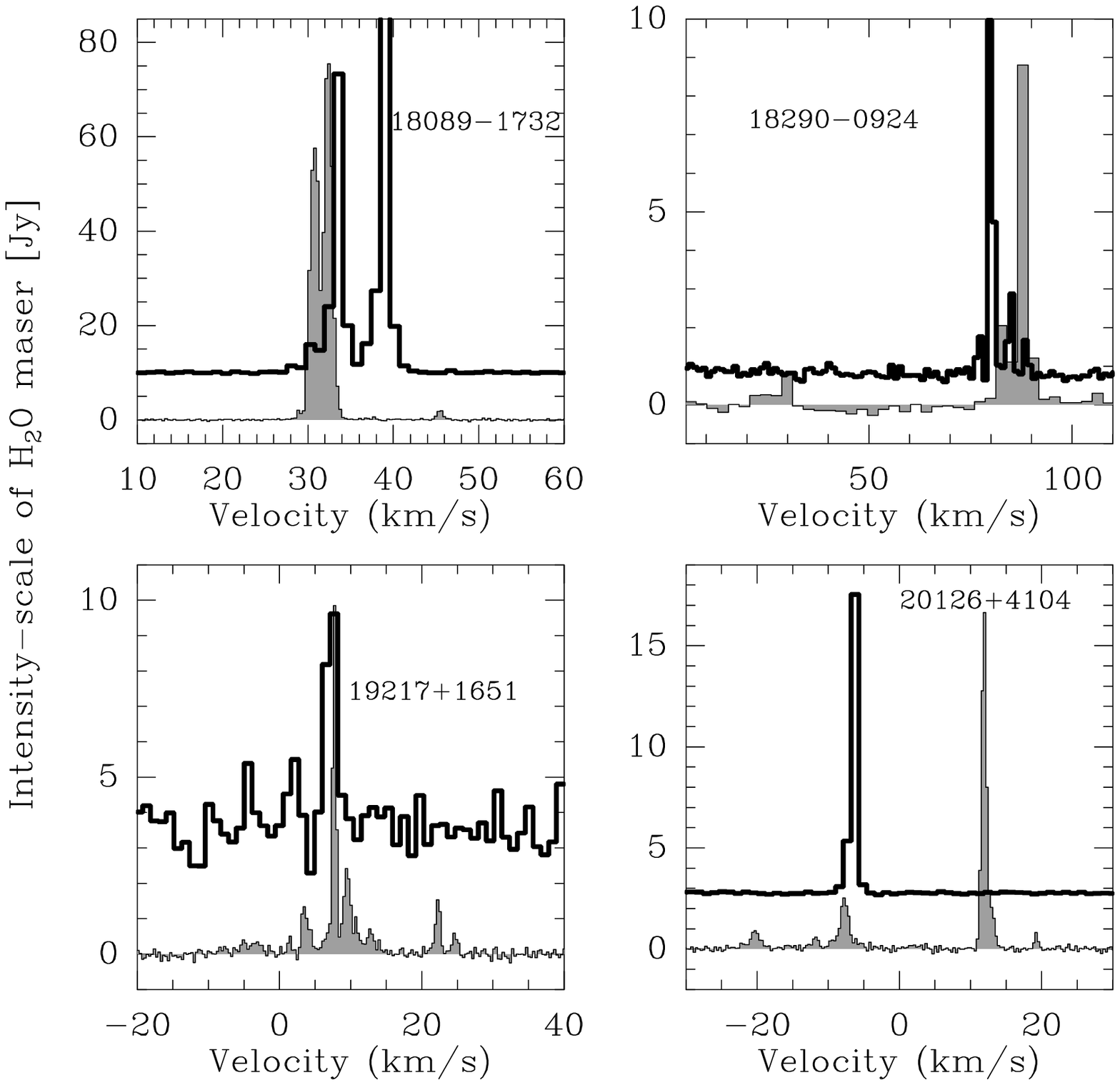} 
 \figcaption[maser_examples.ps]{Water (grey shaded)
  and methanol (thick lines) maser example spectra. The abscissa shows
  the velocities (LSR) for both spectra, while the ordinate shows the
  intensity of the water maser in [Jy], the methanol maser intensity
  is shown in arbitrary units. \label{maserexamples}}

\newpage 
  \includegraphics[angle=-90,width=8cm]{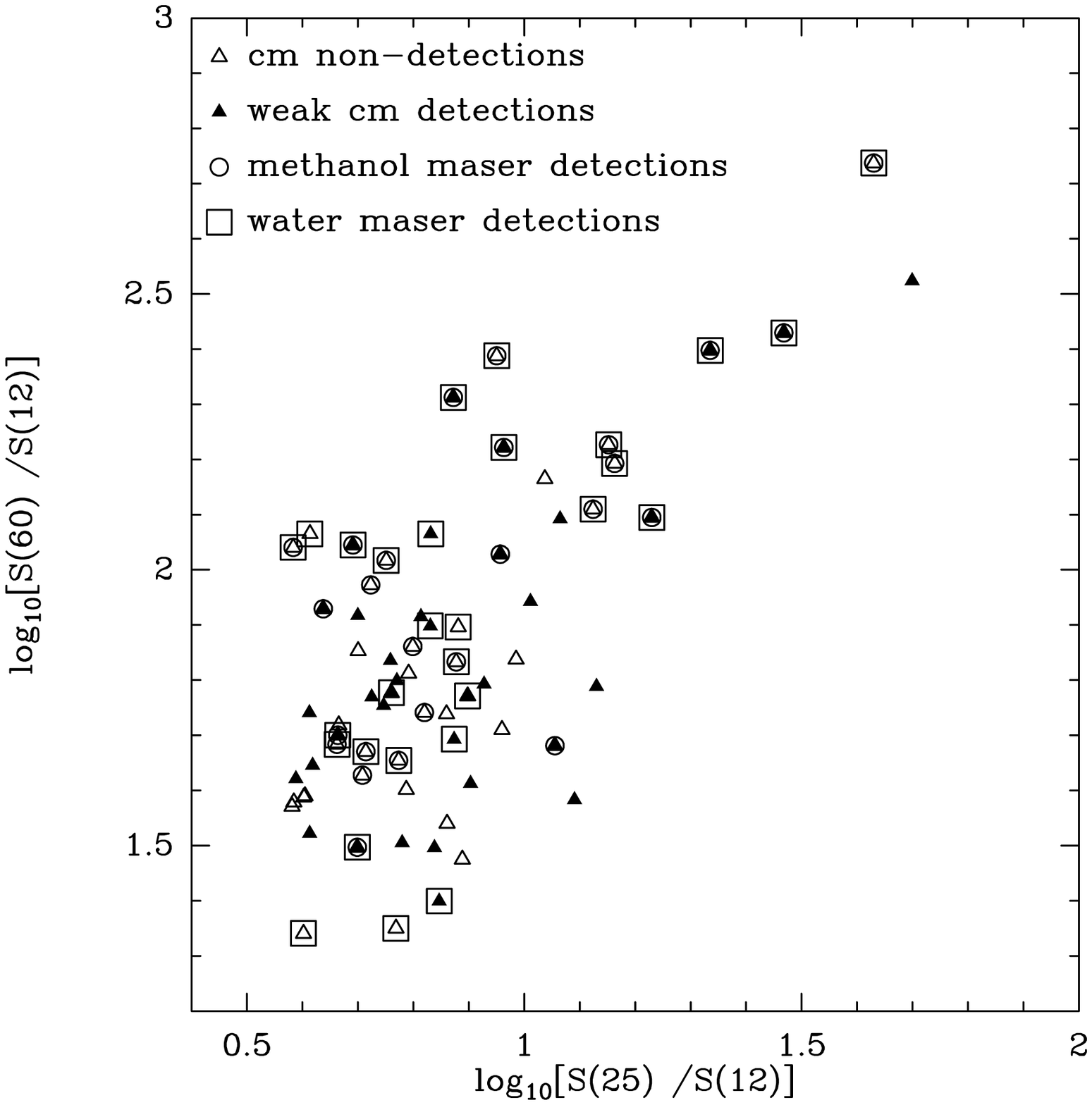}
  \figcaption[sample.ps]{{\it IRAS} Color-color diagram of the whole sample: 
  open
  triangles represent sources without and filled triangles with cm
  emission. The surrounding circles and squares represent detected
  methanol and water maser emission, respectively \label{maserwc}}
  
\includegraphics[angle=-90,width=8cm]{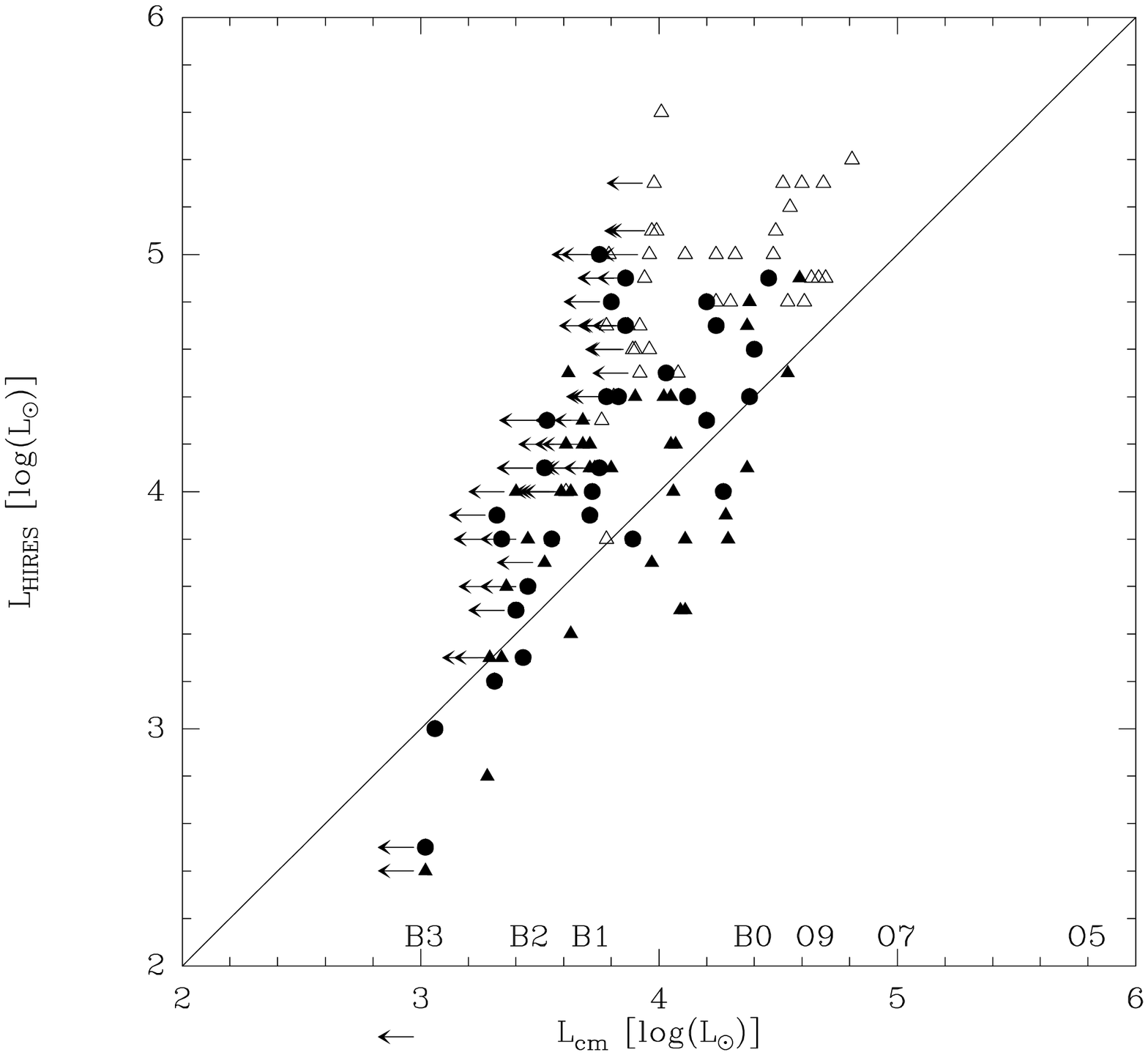}
  \figcaption[lir_lcm.ps]{The HIRES IR luminosities are
  plotted against the luminosities based on the cm observations. Again
  filled circles show sources with no distance ambiguities and
  triangles sources with distance ambiguities (filled: near distance;
  open: far distance) The straight line shows the 1:1
  correlation. \label{luminosities}} 


\begin{thebibliography}{}
   
\bibitem[Aumann et al.\ (1990)]{aumann 1990} Aumann H.H., Fowler J.W., Melnyk M., 1990, AJ, 99, 1674
\bibitem[Bernasconi \& Maeder (1996)]{bernasconi 1996} Bernasconi P.A., Maeder A., 1996, A\&A, 307, 829
\bibitem[Beuther et al.\ (2000)]{beuther 2000} Beuther H., Sridharan T.K., Schilke P., Wyrowski F., Menten K.M., 2000, in: Star Formation from The Small to the Large Scale, ESA SP-445
\bibitem[Beuther et al.\ (2001)]{beuther 2001} Beuther H., Schilke P., Menten K.M., Sridharan T.K., Wyrowski F., Motte F., 2001, ApJ, submitted
\bibitem[Blaauw (1964)]{blaauw 1964} Blaauw A., 1964, Annual Review of A\&A, p. 213
\bibitem[Brand \& Blitz (1993)]{brand 1993} Brand J., Blitz L., 1993, A\&A, 275, 67
\bibitem[Brand et al.\ (1994)]{brand 1994} Brand J., Cesaroni R., Caselli, P., et al.\, 1994, A\&AS, 103, 541
\bibitem[Bronfman et al.\ (1996)]{bronfman 1996} Bronfman L., Nyman L.A., May J., 1996, A\&AS, 115, 81
\bibitem[Bronfman et al.\ (2000)]{bronfman 2000} Bronfman L., Casassus S., May J., Nyman L.-A, 2000, A\&A, 358, 521
\bibitem[Caswell \& Vaile (1995)]{caswell 1995} Caswell J.L., Vaile R.A., 1995, MNRAS, 273, 328
\bibitem[Casoli et al.\ (1986)]{casoli 1986} Casoli F.; Combes F., Dupraz C., Gerin M., Boulanger F., 1986, A\&A, 169, 281
\bibitem[Casassus et al.\ (2000)]{casassus 2000} Casassus S., Bronfman L., May J., Nyman L.-A., 2000, A\&A, 358, 514
\bibitem[Cesaroni et al.\ (1994)]{cesaroni 1994} Cesaroni,R., Churchwell, E., Hofner, P., Walmley, C.M., Kurtz, S., 1994a, A\&A, 252, 278
\bibitem[Cesaroni et al.\ (1997)]{cesaroni 1997} Cesaroni, R., Felli, M., Testi, L., Walmley, C.M., Olmi, L., 1997, A\&A 325, 725
\bibitem[Cesaroni et al.\ (1999)]{cesaroni 1999} Cesaroni R., Felli M., Jenness T., et al.\, 1999, A\&A, 345, 949
\bibitem[Churchwell et al.\ (1990)]{churchwell 1990} Churchwell E., Walmsley C., Cesaroni R., 1990, A\&ASS, 83, 119
\bibitem[Churchwell (2000)]{churchwell 2000} Churchwell E., 2000, in: Unsolved Problems in Stellar Evolution, Space Science Library
\bibitem[Dame et al.\ (1987)]{dame 1987} Dame T.M. Ungerechts H., Cohen R.S., 1987, ApJ, 322, 706
\bibitem[Danby et al.\ (1988)]{danby 1988} Danby G., Flower D.R., Valiron P., Schilke P., Walmsley C.M., 1988, MNRAS, 235, 229
\bibitem[Egan et al.\ (1998)]{egan 1998} Egan M.P., Shipman R.F., Price S.D., Carey S.J., Clark F.O., 1998, ApJ, 494, L199
\bibitem[Fomalont et al.\ (1991)]{fomalont 1991} Fomalont E.B., Windhorst R.A., Fristian J.A., Kellerman K.I., 1991, AJ, 102, 125
\bibitem[Garay et al.\ (1993)]{garay 1993} Garay G., Rodriguez L.F., Moran J.M., Churchwell E, 1993, ApJ, 418, 368
\bibitem[Garay et al.\ (1996)]{garay 1996} Garay G., Ramirez S., Rodriguez L.F., Curiel S., Torrelles J.M., 1996, AJ, 459, 193
\bibitem[Gezari (1982)]{gezari 1982} Gezari D.Y., 1982, ApJ, 259, 29
\bibitem[Gregory \& Condon (1991)]{parkes} Gregory P.C., Condon J.J., 1991, ApJS, 75, 1011
\bibitem[Griffith et al.\ (1994)]{griffith 1994} Griffith M.R., Wright A.E., Burke B.F., Ekers R.D., 1994, ApJS, 90, 179
\bibitem[Henning et al.\ (2000)]{henning 2000} Henning T., Klein R., Launhardt R., Schreyer K., Stecklum B., 2000, in: Infrared Surveys, Springer Verlag, eds. Lemke D.
\bibitem[Hillenbrand \& Hartmann (1998)]{hillenbrand 1998} Hillenbrand L., Hartmann L.W., 1998, ApJ, 492, 540
\bibitem[Hofner et al.\ (1999)]{hofner 1999} Hofner P., Cesaroni R., Rodriguez L.F., Marti J., 1999, A\&A, 345, L43
\bibitem[Hunter (1997)]{hunter 1997} Hunter T., 1997, Ph.D. Thesis, Caltech
\bibitem[Hunter (1998)]{hunter 1998} Hunter T., Neugebauer G., Benford D.J., 
Matthews K., Lis D.C., Serabyn E., Phillips T.G., 1998, ApJ, 493, L97
\bibitem[Hunter et al.\ (2000)]{hunter 2000} Hunter T., Curchwell E., Watson C., Cox P., Benford D.J., Roelfsema P.R., 2000, ApJ, accepted
\bibitem[Jijina et al.\ (1999)]{jijina 1999} Jijina J., Myers P.C., Adams F.C., 1999, ApJS, 125, 161
\bibitem[Kreysa et al.\ (1998)]{kreysa 1998} Kreysa E., Gem\"und H.P., Gromke J., et al.\ 1998, Proc. SPIE, 3357, 319
\bibitem[Kuchar \& Bania (1994)]{kuchar 1994} Kuchar T.A., Bania T.M., 1994, ApJ, 436, 117
\bibitem[Kurtz et al.\ (1994)]{kurtz 1994} Kurtz S., Churchwell E., Wood D.O.S., 1994, ApJ, 91, 1994
\bibitem[Kurtz et al.\ (2000)]{kurtz 2000} Kurtz S., Cesaroni R., Churchwell E., Walmsley C.M., 2000, in Protostars \& Planets IV, ed. V. Mannings
\bibitem[Lada (1993)]{lada 1993} Lada C., 1993, in: The Physics of Star Formation and Early Stellar Evolution, NATO ASI Series, Vol. 342, eds. Lada C., and Kylafis D.
\bibitem[Lang (1992)]{lang 1992} Lang K., 1992, Astrophysical Data, Springer Verlag New York, Inc.
\bibitem[Lis et al.\ (1998)]{lis 1998} Lis D.C., Serabyn E., Keene J., Dowell C.D., Benford D.J., Phillips T.G., 1998, ApJ, 509, 299
\bibitem[Marti et al.\ (1999)]{marti 1999} Marti J., Rodriguez L.F., Torrelles J.M., 1999, A\&A, 345, L5
\bibitem[MacLeod \& Gaylard (1992)]{macleod 1992} MacLeod G.C., Gaylard M.J., 1992, MNRAS, 256, 519
\bibitem[Mathis (1990)]{mathis 1990} Mathis J.S., 1990, ARAA, pp. 37
\bibitem[Massey (1998)]{massey 1998} Massey P., 1998, in: The Stellar Initial Mass function, ASP Conference Series, Vol. 142, eds Gilmore G. and Howell D.
\bibitem[McCluskey (1972)]{mccluskey 1972} McCluskey G.E., Kondo Y, 1972, Astrophysics and Space Science 17, pp. 134-149
\bibitem[Megeath \& Tieftrunk (1999)]{megeath 1999} Megeath S.T., Tieftrunk A.R., 1999, ApJ, 526, L113
\bibitem[Menten \& Reid (1995)]{menten 1995} Menten K.M., Reid M.J., 1995, ApJ, 445, L157
\bibitem[Menten (1996)]{menten 1996} Menten K.M., 1996, in IAU Symposium  178, ed. E. v. Dishoek
\bibitem[Menten et al.\ (1999)]{menten 1999} Menten K.M., Sridharan T.K. Wyrowski F., Schilke P., 1999, in: The Physics and Chemistry of the Interstellar Medium, ed. V. Ossenkopf, GCA-Verlag Herdecke 
\bibitem[Mezger et al.\ (1990)]{mezger 1990} Mezger P.G., Wink J.E., Zylka R., 1990, A\&A, 228, 95
\bibitem[Molinari et al.\ (1996)]{molinari 1996} Molinari S., Brand J., Cesaroni R., Palla F., 1996, A\&A, 308, 573
\bibitem[Molinari et al.\ (1998a)]{molinari 1998a} Molinari S., Testi L., Brand J., Cesaroni R., Palla F., 1998, A\&A, 505, L39
\bibitem[Molinari et al.\ (1998b)]{molinari 1998b} Molinari S., Brand J., Cesaroni R., Palla F., Palumbo G., 1998, A\&A, 336, 339 
\bibitem[Molinari et al.\ (2000)]{molinari 2000} Molinari S., Brand J., Cesaroni R., Palla F., 2000, A\&A, 355, 617
\bibitem[Norberg \& Maeder (2000)]{norberg 2000} Norberg P., Maeder A., 2000, A\&A, 359, 1025 
\bibitem[Osorio et al.\ (1999)]{osorio 1999} Osorio M., Lizano S., D'Alessio P., 1999, ApJ, 525, 808
\bibitem[Palla \& Stahler (1993)]{palla 1993} Palla F., Stahler S., 1993, ApJ, 418, 414
\bibitem[Porras et al.\ (2000)]{porras 2000} Porras A., Cruz-Gonzales I., Salas L., 2000, A\&A, in press
\bibitem[Ramesh \& Sridharan (1997)]{rs} Ramesh B., Sridharan T.K., 1997, MNRAS, 284, 1001,
\bibitem[Reid et al.\ (1995)]{reid 1995} Reid M.J., Argon A.L., Masson C.R., Menten K.M., Moran J.M., 1995, ApJ, 443, 238
\bibitem[Sanders et al.\ (1986)]{sanders 1986} Sanders D.B., Clemens D.P., Scoville N.Z., Solomon P.M., 1986, ApJS, 60, 1
\bibitem[Scalo (1986)]{scalo 1986} Scalo J., 1986, Fund. Cosmic Phys., 11, 1
\bibitem[Schatzman \& Praderie (1993)]{schatzman 1993} Schatzman E., Praderie F., 1993, The Stars, Springer Verlag
\bibitem[Schilke et al.\ (1997)]{schilke 1997} Schilke P., Walmsley C., Pineau de For\^ets, Flower D., 1997, A\&A, 321, 293
\bibitem[Shepherd et al.\ (1996)]{shepherd 1996} Shepherd D.S., Churchwell E., 1996, ApJ, 472, 225 
\bibitem[Shepherd et al.\ (1998)]{shepherd 1998} Shepherd D.S., Watson A., Sargent A., Churchwell E., 1998, ApJ, 507, 861
\bibitem[Shepherd et al.\ (2000)]{shepherd 2000} Shepherd D.S., Yu K.C., Bally J., Testi L., 2000, ApJ, 535, 833
\bibitem[Slysh et al.\ (1999)]{slysh 1999} Slysh V.I., Val'tts I.E., Kalenski S.V., Voronkov M.A., Palagi F., Tofani G., Catarzi M., 1999, A\&AS, 134, 115
\bibitem[Snell et al.\ (1990)]{snell 1990} Snell R.L., Dickman R.L., Huang Y.-L., 1990, ApJ, 352, 139
\bibitem[Sridharan et al.\ (1999)]{sridharan 1999} Sridharan T.K., Menten K.M., Wyrowski F., Schilke P., 1999, in: Star Formation 1999, ed. T. Nakamato, The Nobeyama Radio Observatory
\bibitem[Stahler et al.\ (2000)]{stahler 2000} Stahler S., Palla F., Ho P., 2000, in Protostars \& Planets IV, The University of Arizona Press
\bibitem[Szymczak et al.\ (2000)]{szymczak 2000} Szymczak M., Hrynek G., Kus A., 2000, A\&AS, 143, 269
\bibitem[Testi (2000)]{testi 2000} Testi L., talk held at the workshop on massive star formation, Volterra, June 2000
\bibitem[Tieftrunk et al.\ (1997)]{tieftrunk 1997} Tieftrunk A.R., Gaume R.A., Claussen M.J., Wilson T.L., Johnston K.J., 1997, A\&A, 318, 931
\bibitem[Ungerechts et al.\ (1986)]{ungerechts 1986} Ungerechts H., Walmsley M.C., Winnewisser G., 1986, A\&A, 157, 207
\bibitem[Walmsley (1995)]{walmsley 1995} Walmsley M.C., 1995, RevMexAA, 1, 137
\bibitem[Walsh et al.\ (1997)]{walsh 1997} Walsh A.J., Hyland A.R., Robinson G., Burton M.G., 1998, MNRAS, 291, 261
\bibitem[Walsh et al.\ (1998)]{walsh 1998} Walsh A.J., Burton M.G., Hyland A.R., Robinson G., 1998, MNRAS, 301, 640
\bibitem[Walsh et al.\ (2001)]{walsh 2001} Walsh A.J., Bertoldi F., Burton M.G., Nikola T., 2001, MNRAS. in press
\bibitem[Wilking et al.\ (1989)]{wilking 1989} Wilking B.A., Mundy L.G., Blackwell J.H., Howe J.E., 1989, ApJ, 345, 257
\bibitem[van der Walt et al.\ (1995)]{walt 1995} van der Walt D.J., Gaylard M., MacLeod G.C., 1995, A\&AS, 110, 81
\bibitem[Wilner et al.\ (1996)]{wilner 1996} Wilner, D.J., Welch, W.J., Forster, J.R., 1995, ApJ, 449, L73
\bibitem[Wolfire \& Cassinelli (1987)]{wolfire 1987} Wolfire M.G. \& Cassinelli J.P., 1987, ApJ, 319, 850
\bibitem[Wood \& Churchwell (1989a)]{wc89a} Wood, D.O.S., Churchwell, E., 1989, ApJ, 340, 265
\bibitem[Wood \& Churchwell (1989b)]{wc89b} Wood, D.O.S., Churchwell, E., 1989, ApJS, 69, 831
\bibitem[Wright et al.\ (1994)]{wright 1994} Wright A.E., Griffith M.R., Burke B.F., Ekers R.D., 1994, ApJS, 91, 111
\bibitem[Wyrowski et al.\ (1999)]{wyrowski 1999} Wyrowski F., Schilke P., Walmsley C.M.; Menten K.M., 1999, ApJ, 514, L43
\bibitem[Zhang et al.\ (1999)]{zhang 1999} Zhang Q., Hunter T.R., Sridharan T.K., Cesaroni R., 1999, ApJ, 527, L117
\bibitem[Zhang et al.\ (2001)]{zhang 2001} Zhang Q., Hunter T.R., Brand J., 
Sridharan T.K., Molinari S., Kramer M., Cesaroni R., 2001, ApJ, accepted
\bibitem[Zylka (1998)]{zylka 1998} Zylka R., 1998, Pocket Cookbook for the MOPSI Software, MPIfR internal report

\end{thebibliography}
\end{document}